\documentclass[11pt,draftcls, onecolumn]{IEEEtran}

\usepackage{amsmath}
\usepackage{amssymb}
\usepackage[dvips]{graphicx}
\usepackage{epsfig}

\newcommand{\K}{{\sf{K}}}
\newcommand{\x}{\mathbf{x}}
\newcommand{\y}{\mathbf{y}}

\newcommand{\z}{\mathbf{z}}

\newcommand{\n}{\mathbf{n}}

\newcommand{\I}{\mathbf{I}}

\newcommand{\C}{{\sf{C}}}
\newcommand{\bv}{\mathbf{v}}
\newcommand{\bw}{\mathbf{w}}
\newcommand{\dd}{\dagger}

\newcommand{\he}{\hat{h}}
\newcommand{\hn}{\tilde{h}}
\newcommand{\CN}{\mathcal{CN}}
\newcommand{\tgamma}{\tilde{\gamma}}
\newcommand{\D}{\mathcal{D}}

\newcommand{\figsize}{0.7}
\newcommand{\ffigsize}{0.7}

\newcommand{\tsnr}{{\text{\footnotesize{SNR}}}}

\newcommand{\ud}{{\mathrm{d}}}

\newtheorem{lemma:bitenergylow}{Proposition}
\newtheorem{lemma:bitenergyhigh}[lemma:bitenergylow]{Lemma}
\newtheorem{lemma:differentiation}{Differentiation Lemma}
\newtheorem{prop:flashminbitenergy}[lemma:bitenergylow]{Proposition}
\newtheorem{prop:pasympcap}[lemma:bitenergylow]{Proposition}

\newtheorem{prop:asympcap}{Theorem}

\newtheorem{prop:flashbitenergy}[prop:asympcap]{Theorem}

\newtheorem{prop:blisotropic}[prop:asympcap]{Theorem}
\newtheorem{prop:eqvmutual}[prop:asympcap]{Theorem}
\newtheorem{prop:existence+ktc}[prop:asympcap]{Theorem}
\newtheorem{theo:discrete}[prop:asympcap]{Theorem}
\newtheorem{theo:bitenergysinglemass}[prop:asympcap]{Theorem}
\newtheorem{theo:discreteinterl}[prop:asympcap]{Theorem}
\newtheorem{theo:bitenergyook}[prop:asympcap]{Theorem}

\begin{document}

\title{On the Capacity and Energy Efficiency of Training-Based Transmissions over Fading Channels }



%
\author{\authorblockN{Mustafa Cenk Gursoy}
\\ \vspace{-.1cm}
\authorblockA{Department of Electrical Engineering\\ \vspace{-.1cm}
University of Nebraska-Lincoln, Lincoln, NE 68588\\
\vspace{-.1cm}Email: gursoy@engr.unl.edu}}


\maketitle \vspace{-2cm}
\begin{abstract}\footnote{This work was supported in part by the NSF CAREER Grant
CCF-0546384. The material in this paper was presented in part at the
IEEE International Symposium on Information Theory (ISIT), Nice,
France, June 2007.} In this paper, the capacity and energy
efficiency of training-based communication schemes employed for
transmission over a-priori unknown Rayleigh block fading channels
are studied. In these schemes, periodically transmitted training
symbols are used at the receiver to obtain the minimum
mean-square-error (MMSE) estimate of the channel fading
coefficients. Initially, the case in which the product of the
estimate error and transmitted signal is assumed to be Gaussian
noise is considered. In this case, it is shown that bit energy
requirements grow without bound as the signal-to-noise ratio
($\tsnr$) goes to zero, and the minimum bit energy is achieved at a
nonzero $\tsnr$ value below which one should not operate. The effect
of the block length on both the minimum bit energy and the $\tsnr$
value at which the minimum is achieved is investigated. Energy
efficiency analysis is also carried out when peak power constraints
are imposed on pilot signals. Flash training and transmission
schemes are analyzed and shown to improve the energy efficiency in
the low-$\tsnr$ regime.

In the second part of the paper, the capacity and energy efficiency
of training-based schemes are investigated when the channel input is
subject to peak power constraints. The capacity-achieving input
structure is characterized and the magnitude distribution of the
optimal input is shown to be discrete with a finite number of mass
points. The capacity, bit energy requirements, and optimal resource
allocation strategies are obtained through numerical analysis. The
bit energy is again shown to grow without bound as $\tsnr$ decreases
to zero due to the presence of peakedness constraints. Capacity and
energy-per-bit are also analyzed under the assumptions that the
transmitter interleaves the data symbols before transmission over
the channel, and per-symbol peak power constraints are imposed. The
improvements in energy efficiency when on-off keying with fixed peak
power and vanishing duty cycle is employed are studied. Comparisons
of the performances of training-based and noncoherent transmission
schemes are provided.


\end{abstract}

\section{Introduction}

In wireless communications, channel conditions vary randomly over
time due to mobility and changing environment, and the degree of
channel side information (CSI) assumed to be available at the
receiver and transmitter is a key assumption in the study of
wireless fading channels. The case in which the channel is assumed
to be perfectly known at the receiver and/or transmitter has been
extensively studied. In an early work, Ericsson \cite{Ericsson}
obtained the capacity of flat fading channels with perfect receiver
CSI. More recently, Ozarow \emph{et al.} \cite{Ozarow} studied the
average and outage capacity values in the cellular mobile radio
setting assuming perfect channel knowledge at the receiver.
Goldsmith and Varaiya \cite{Goldsmith Varaiya} analyzed the capacity
of flat fading channels with perfect CSI at the transmitter and/or
receiver.

The assumption of having perfect channel knowledge is unwarranted
when communication is trying to be established in a highly mobile
environment. This consideration has led to another line of work
where both the receiver and transmitter are assumed to be completely
uninformed of the channel conditions. Abou-Faycal \emph{et al.}
\cite{Abou} studied the capacity of the unknown Rayleigh fading
channel and showed that the optimal input amplitude has a discrete
structure. This is in stark contrast to the optimality of a
continuous Gaussian input in known channels. In \cite{Gursoy-part1}
and \cite{gursoy-isit}, the discreteness of the capacity-achieving
amplitude distribution is proven for noncoherent Rician fading
channels under input peakedness constraints. When the input is
subject to peak power constraints, the discrete nature of the
optimal input is shown for a general class of single-input
single-output channels in \cite{Meyn}. Marzetta and Hochwald
\cite{Marzetta} gave a characterization of the optimal input
structure for unknown multiple-antenna Rayleigh fading channels.
This analysis subsequently led to the proposal of unitary space-time
modulation techniques \cite{Marzetta Hochwald2}. Chan \emph{et al}
\cite{Chan} considered conditionally Gaussian multiple-input
multiple-output (MIMO) channels with bounded inputs and proved the
discreteness of the optimal input under certain conditions. Zheng
and Tse \cite{Zheng1} analyzed the multiple-antenna Rayleigh
channels and identified the high signal-to-noise ratio (SNR)
behavior of the channel capacity.

Heretofore, the two extreme assumptions of having either perfect CSI
or no CSI have been discussed. Practical wireless systems live in
between these two extremes. Unless there is very high mobility,
wireless systems generally employ estimation techniques to learn the
channel conditions, albeit with errors. Hence, it is of utmost
interest to analyze fading channels with imperfect CSI.
M\'edard \cite{Medard} investigated the effect upon channel capacity
of imperfect channel knowledge and obtained upper and lower bounds
on the input-output mutual information. Lapidoth and Shamai
\cite{Lapidoth Shamai} analyzed the effects of channel estimation
errors on the performance if Gaussian codebooks are used and nearest
neighbor decoding is employed. The capacity of imperfectly-known
fading channels is characterized in the low-$\tsnr$ regime in
\cite{Verdu} and in the high-$\tsnr$ regime in \cite{Lapidoth
Moser}.

The aforementioned studies have not considered explicit training and
estimation techniques, and resources allocated to them. Recently,
Hassibi and Hochwald \cite{Hassibi} studied training schemes to
learn the multiple-antenna channels. In this work, power and time
dedicated to training is optimized by maximizing a lower bound on
the capacity. Similar training techniques are also discussed in
\cite{Zheng1}. Due to its practical significance, the
information-theoretic analysis of training schemes has attracted
much interest (see e.g., \cite{Baltersee1}-\cite{sami2}). Since
exact capacity expressions are difficult to find,  these studies
have optimized the training signal power, duration, and placement
using capacity bounds. Since Gaussian noise is the worst-case
uncorrelated additive noise in a Gaussian setting \cite{Hassibi}, a
capacity lower bound is generally obtained by assuming the product
of the estimate error and the transmitted signal as another source
of Gaussian noise. In the above cited work, training symbols are
employed to solely facilitate channel estimation. However, we note
that training symbols can also be used for timing- and
frequency-offset synchronization, and channel equalization
\cite{Gansman}-\cite{Kaleh}. Tong \emph{et al.} in \cite{Tong}
present an overview of pilot-assisted wireless transmissions and
discuss design issues from both information-theoretic and signal
processing perspectives.


Another important concern in wireless communications is the
efficient use of limited energy resources. In systems where energy
is at a premium, minimizing the energy cost per unit transmitted
information will improve the efficiency. Hence, the energy required
to reliably send one bit is a metric that can be adopted to measure
the performance. Generally, energy-per-bit requirement is minimized,
and hence the energy efficiency is maximized, if the system operates
in the low-$\tsnr$ regime. In \cite{Verdu}, Verd\'u has analyzed the
tradeoff between the spectral efficiency and bit energy in the
low-$\tsnr$ regime for a general class of channels and shown that
the normalized received minimum bit energy of $-1.59$ dB is achieved
as $\tsnr \to 0$ in averaged power limits channels regardless of the
availability of CSI at the receiver. On the other hand, \cite{Verdu}
has proven that if the receiver has imperfect CSI, the wideband
slope, which is the slope of the spectral efficiency curve at zero
spectral efficiency, is zero. Hence, approaching the minimum bit
energy of $-1.59$ dB is extremely slow, and moreover it requires
input signals with increasingly higher peak-to-average power ratios.
The impact upon the energy efficiency of limiting the peakedness of
signals is analyzed in \cite{Gursoy-part2}. The wideband channel
capacity in the presence of input peakedness constraints is
investigated in \cite{Medard Gallager}, \cite{Subramanian Hajek},
and \cite{Telatar Tse}.

Energy efficiency, which is of paramount importance in many wireless
systems, has not been the core focus of the aforementioned work on
training schemes. Moreover, previous studies optimized the training
parameters by using capacity lower bounds. These achievable rate
expressions are relevant for systems in which the channel estimate
is assumed to be perfect and transmission and reception is designed
for a known channel. Note that these assumptions will lead to poor
performance unless the $\tsnr$ is high or the channel coherence time
is long.

The contributions of this paper are the following:

\begin{itemize}

\item We provide an energy efficiency perspective
by analyzing the performance of training techniques in the
low-$\tsnr$ regime. Note that at low $\tsnr$ levels, the quality of
the channel estimate is far from being perfect. We quantify the
performance losses in terms of energy efficiency in the worst-case
scenario where the estimate is assumed to be perfect. We identify an
$\tsnr$ level below which one should avoid operating. We consider
flash training and transmission techniques to improve the
performance.

\item We obtain the exact capacity of training-based schemes by
characterizing the structure of the capacity-achieving input
distribution under input peak power constraints which are highly
relevant in practical applications. Optimal resource allocation is
performed using the exact capacity values. Improvements in energy
efficiency with respect to the worst-case scenario are shown.

\item We compare the performances of untrained noncoherent
and training-based communication schemes under peak power
constraints and show through numerical results that performance loss
experienced by training-based schemes is small even at low $\tsnr$
levels and small values of coherence time. On the other hand, if
data symbols are interleaved and experience independent fading, we
show that training-based schemes outperform noncoherent techniques.

\item We find the attainable bit energy levels in the low-$\tsnr$ regime
when limitations on the peak-to-average power ratio are relaxed and
on-off keying with fixed power and vanishing duty cycle is used to
transmit information.
\end{itemize}

The organization of the paper is as follows. Section
\ref{sec:channelmodel} provides the channel model. In Section
\ref{sec:trainingscheme}, training-based transmission and reception
is described. In Section \ref{sec:worstcase}, we study the
achievable rates and energy efficiency in the case where the product
of the channel estimate and the transmitted signal is assumed to be
Gaussian noise. In Section \ref{sec:peakpower}, we analyze the
capacity and the energy efficiency of training-based schemes when
the input is subject to peak power limitations. Section
\ref{sec:conclusion} includes our conclusions. Proofs of several
results are relegated to the Appendix.

\section{Channel Model} \label{sec:channelmodel}

We consider Rayleigh block-fading channels where the input-output
relationship within a block of $m$ symbols is given by
\vspace{-.2cm}
\begin{gather}\label{eq:model}
\y = h \x + \n 
\end{gather}
where $h \sim \mathcal{CN}(0,\gamma^2)$ \footnote{$\x \sim
\mathcal{CN}(\mathbf{d}, \mathbf{\Sigma})$ is used to denote that
$\x$ is a complex Gaussian random vector with mean $E\{\x\} =
\mathbf{d}$ and covariance $E\{(\x - \mathbf{d})(\x -
\mathbf{d})^\dd\} = \mathbf{\Sigma}$} is a zero-mean circularly
symmetric complex Gaussian random variable with variance $E\{|h|^2\}
= \gamma^2$, and $\n$ is a zero-mean, $m$ complex-dimensional
Gaussian random vector\footnote{Note that in the channel model
(\ref{eq:model}), $\y$, $\x$, and $\n$ are column vectors.} with
covariance matrix $E\{\n \n^\dd\} = N_0 \I$. $\x$ and $\y$ are the
$m$ complex-dimensional channel input and output vectors,
respectively. It is assumed that the fading coefficients stay
constant for a block of $m$ symbols and have independent
realizations for each block. It is further assumed that neither the
transmitter nor the receiver has prior knowledge of the realizations
of the fading coefficients.

\section{Training-Based Transmission and Reception}
\label{sec:trainingscheme}

We assume that pilot symbols are employed in the system to
facilitate channel estimation at the receiver. Hence, the system
operates in two phases, namely training and data transmission. In
the training phase, pilot symbols known at the receiver are sent
from the transmitter and the received signal is
\begin{gather}
\y_t = h \x_t + \n_t
\end{gather}
where $\y_t$, $\x_t$, and $\n_t$ are $l$-dimensional vectors
signifying the fact that $l$ out of $m$ input symbols are devoted to
training. It is assumed that the receiver employs minimum
mean-square error (MMSE) estimation to obtain the estimate
\begin{gather}
\he = E\{h | \y_t\} = \frac{\gamma^2}{\gamma^2 \|\x_t\|^2 + N_0}
\x_t^\dd \y_t.
\end{gather}
With this estimate, the fading coefficient can now be expressed as
\begin{gather}
h = \he + \hn
\end{gather}
where
\begin{align}
\he \sim \CN \left( 0, \frac{\gamma^4 \|\x_t\|^2}{\gamma^2
\|\x_t\|^2 + N_0}\right) \quad \text{and} \quad \hn \sim \CN \left(
0, \frac{\gamma^2 N_0 }{\gamma^2 \|\x_t\|^2 + N_0}\right).
\label{eq:he}
\end{align}
Note that $\hn$ denotes the error in the channel estimate. Following
the training phase, the transmitter sends the ($m-l$)-dimensional
data vector $\x_d$, and the receiver equipped with the knowledge of
the channel estimate operates on the received signal
\begin{gather} \label{eq:dataphasemodel}
\y_d = \he \x_d + \hn \x_d + \n_d
\end{gather}
to recover the transmitted information. We note that since
training-based schemes are studied in this paper, memoryless fading
channels in which $m = 1$ are not considered, and it is assumed
throughout the paper that the block length satisfies $m \ge 2$.

\section{Achievable Rates and Energy Efficiency in the Worst Case
Scenario} \label{sec:worstcase}

\subsection{Average Power Limited Case}
\label{subsec:worstcaseavgpower}

In this section, we assume that the input is subject to an average
power constraint
\begin{gather} \label{eq:avgpower}
E\{\|\x\|^2\} \le mP. 
\end{gather}
Our overall goal is to identify the bit energy values that can be
attained with optimized training parameters such as the power and
duration of pilot symbols. The least amount of energy required to
send one information bit reliably is given by\footnote{Note that
$\frac{E_b}{N_0}$ is the bit energy normalized by the noise power
spectral level $N_0$.}
\begin{gather}
\frac{E_b}{N_0} = \frac{\tsnr}{C(\tsnr)}
\end{gather}
where $C(\tsnr)$ is the channel capacity in bits/symbol. In this
section, we follow the general approach in the literature and
consider a lower bound on the channel capacity by assuming that
\begin{gather}
\z = \hn \x_d + \n_d
\end{gather}
is a Gaussian noise vector that has a covariance of
\begin{gather}
E\{\z \z^\dd\} =\sigma_{\hn}^2 E\{\x_d \x_d^\dd\} + N_0 \I,
\end{gather}
and is uncorrelated with the input signal $\x_d$. With this
assumption, the channel model becomes
\begin{gather} \label{eq:worstcasemodel}
\y_d = \he \x_d + \z.
\end{gather}
This model is called the worst-case scenario since the channel
estimate is assumed to be perfect, and the noise is modeled as
Gaussian, which presents the worst case \cite{Hassibi}. The capacity
of the channel in (\ref{eq:worstcasemodel}), which acts as a lower
bound on the capacity of the channel in (\ref{eq:dataphasemodel}),
is achieved by a Gaussian input with
\begin{gather}
E\{\x_d \x_d^\dd\} = \frac{(1-\delta^*)mP}{m-1} \I
\end{gather}
where
$\delta^*$ is the optimal fraction of power allocated to the pilot
symbol, i.e., $|x_t|^2 = \delta^* mP$. The optimal value is given by
\begin{gather}
\delta^* = \sqrt{\eta(\eta+1)} - \eta
\end{gather}
where
\begin{gather} \label{eq:eta+snr}
\eta = \frac{m \, \tsnr +(m-1)}{m(m-2) \tsnr} \quad \text{and} \quad
\tsnr = \frac{\gamma^2 P}{N_0}.
\end{gather}
Note that $\tsnr$ in (\ref{eq:eta+snr}) is the received
signal-to-noise ratio. In the average power limited case, sending a
single pilot is optimal because instead of increasing the number of
pilot symbols, a single pilot with higher power can be used and a
decrease in the duration of the data transmission can be avoided.
Hence, the optimal $\x_d$ is an ($m-1$)-dimensional Gaussian vector.
Since the above results are indeed special cases of those in
\cite{Hassibi}, the details are omitted. The resulting capacity
expression\footnote{Unless specified otherwise, all logarithms are
to the base $e$.} is
\begin{align}
C_L(\tsnr) &= \frac{m\!-\!\!1}{m} E_w \left\{ \log \left( 1 +\!
\frac{\phi(\tsnr)\tsnr^2}{\psi(\tsnr)\tsnr + (m\!-\!\!1)}
|w|^2\right) \right\} \nonumber
\\
&= \frac{m\!-\!\!1}{m} E_w \left\{ \log \left( 1 + f(\tsnr)
|w|^2\right) \right\} \text{ nats/symbol} \label{eq:worstcap}
\end{align}
where
\begin{gather}
\phi(\tsnr) = \delta^*(1-\delta^*)m^2, \quad \text{and} \quad
\psi(\tsnr) = (1 + (m-2)\delta^*)m,
\end{gather}
and $w \sim \CN(0,1)$. Note also that the expectation in
(\ref{eq:worstcap}) is with respect to the random variable $w$. The
bit energy values in this setting are given by
\begin{gather} \label{eq:worstengy}
\frac{E_{b,U}}{N_0} = \frac{\tsnr}{C_L(\tsnr)} \log2
\end{gather}
where $C_L$ is in nats/symbol. $\frac{E_{b,U}}{N_0}$ provides the
least amount of normalized bit energy values in the worst-case
scenario and also serves as an upper bound on the achievable bit
energy levels of channel (\ref{eq:dataphasemodel}). It is shown in
\cite{Lapidoth Shamai} that if the channel estimate is assumed to be
perfect, and Gaussian codebooks designed for known channels are
used, and scaled nearest neighbor decoding is employed at the
receiver, then the generalized mutual information has an expression
similar to (\ref{eq:worstcap}) (see \cite[Corollary 3.0.1]{Lapidoth
Shamai}). Hence $\frac{E_{b,U}}{N_0}$ also gives a good indication
of the energy requirements of a system operating in this fashion.
The next result provides the asymptotic behavior of the bit energy
as $\tsnr$ decreases to zero.

\begin{lemma:bitenergylow} \label{lemma:bitenergylow}
The normalized bit energy (\ref{eq:worstengy}) grows without bound
as the signal-to-noise ratio decreases to zero, i.e.,
\begin{gather}
\left.\frac{E_{b,U}}{N_0}\right|_{C_L = 0} = \lim_{\tsnr \to 0}
\frac{\tsnr}{C_L(\tsnr)}\log2 = \frac{\log2}{\dot{C}_L(0)} = \infty.
\end{gather}
\end{lemma:bitenergylow}
\vspace{.5cm} \emph{Proof}: In the low SNR regime, we have
\begin{align}
C_L(\tsnr) &= \frac{m-1}{m} \left( f(\tsnr) E\{|w|^2\} +
o(f(\tsnr))\right)
\\
&= \frac{m-1}{m} \left( f(\tsnr) + o(f(\tsnr))\right).
\end{align}
As $\tsnr \to 0$, $\delta^* \to 1/2$, and hence $\phi(\tsnr) \to m^2
/4$ and $\psi(\tsnr) \to m + m(m-2)/2$. Therefore, it can easily be
seen that \vspace{-.2cm}
\begin{gather}
f(\tsnr) = \frac{m^2}{4(m-1)} \tsnr^2 + o(\tsnr^2)
\end{gather}
from which we have $\dot{C}_L(0) = 0$. \hfill $\square$

The fact that $C_L$ decreases as $\tsnr^2$ as $\tsnr$ goes to zero
has already been pointed out in \cite{Hassibi}. The reason for this
behavior is that as $\tsnr$ decreases, the power of $\he$
(\ref{eq:he}) decreases linearly with $\tsnr$ and hence the quality
of the channel estimate deteriorates. Since the channel estimate is
assumed to be perfect, the effective signal-to-noise ratio decays as
$\tsnr^2$ leading to the observed result. Proposition
\ref{lemma:bitenergylow} shows the impact of this behavior on the
energy-per-bit, and indicates that it is extremely
energy-inefficient to operate at very low $\tsnr$ values. The result
holds regardless of the size of the block length $m$ as long as it
is finite.
We further conclude that in a training-based scheme where the
channel estimate is assumed to be perfect, the minimum energy per
bit is achieved at a nonzero $\tsnr$ value. This most
energy-efficient operating point can be obtained by numerical
analysis. We can easily compute $C_L(\tsnr)$ in (\ref{eq:worstcap}),
and hence the bit energy values.
\begin{figure}
\begin{center}
\includegraphics[width = \ffigsize\textwidth]{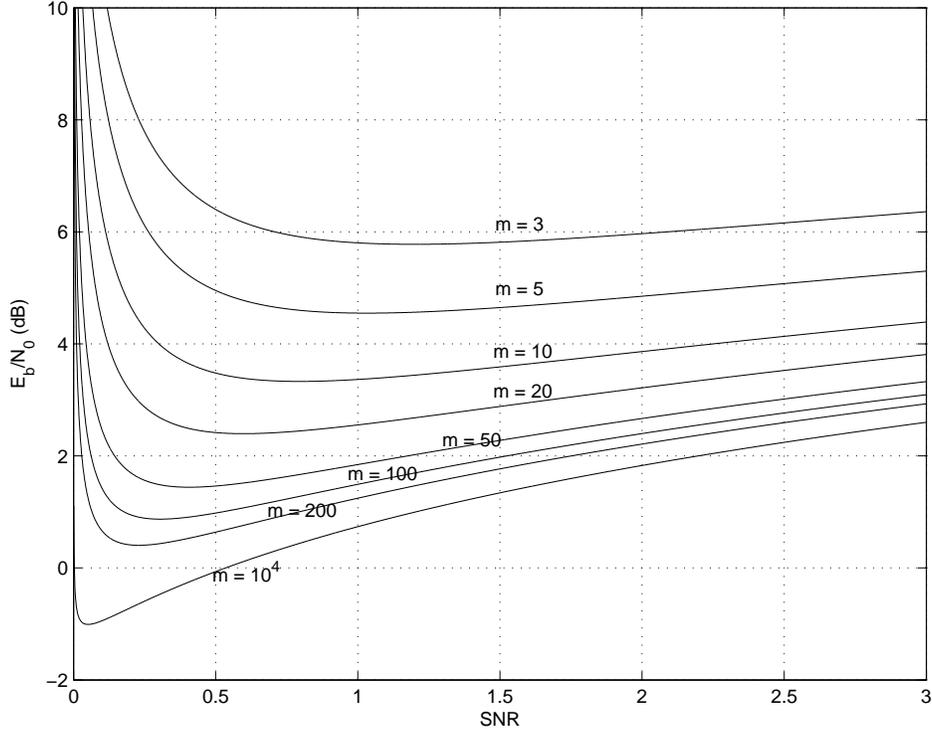}
\caption{Energy per bit $E_{b,U}/N_0$ vs. $\tsnr$ in the worst-case
scenario} \label{fig:trainingbitenergy}
\end{center}
\end{figure}

Figure \ref{fig:trainingbitenergy} plots the normalized bit energy
curves as a function of $\tsnr$ for block lengths of $m = 3, 5, 10,
20, 50, 100$, $200, 10^4$. As predicted, for each block length
value, the minimum bit energy is achieved at nonzero $\tsnr$, and
the bit
energy requirement increases as $\tsnr \to 0$. 
It is been noted in \cite{Hassibi} that training-based schemes,
which assume the channel estimate to be perfect, perform poorly at
very low $\tsnr$ values, and the exact transition point below which
one should not operate in this fashion is deemed as not clear. Here,
we propose the $\tsnr$ level at which the minimum bit energy is
achieved as a transition point since operating below this point
results in higher bit energy requirements. It is further seen in
Fig. \ref{fig:trainingbitenergy} that the minimum bit energy is
attained at an $\tsnr$ value that satisfies
\begin{gather}
\frac{d}{d \tsnr}\left(\frac{E_{b,U}}{N_0}\right) = \frac{d}{d
\tsnr}\left(\frac{\tsnr \log2}{C_L(\tsnr)}\right) = 0.
\end{gather}
Another observation from Fig. \ref{fig:trainingbitenergy} is that
the minimum bit energy decreases with increasing $m$ and is achieved
at a lower $\tsnr$ value. The following result sheds a light on the
asymptotic behavior of the capacity as $m \to \infty$.
\begin{prop:asympcap} \label{prop:asympcap}
As the block length $m$ increases, $C_L$ approaches to the capacity
of the perfectly known channel, i.e.,\vspace{-.1cm}
\begin{gather}
\lim_{m \to \infty} C_L(\tsnr) = E_w\{ \log(1 + \tsnr |w|^2)\}.
\end{gather}
Moreover, define $\chi = 1/m$. Then
\begin{gather}
\left.\frac{dC_{L}(\tsnr)}{d\chi}\right|_{\chi = 0} = -\infty.
\label{eq:derivative}
\end{gather}
\end{prop:asympcap}

\vspace{0cm} \emph{Proof}: We have
\begin{align}
\lim_{m \to \infty} C_L(\tsnr) &= \lim_{m \to \infty} E_w \left\{
\log \left( 1 + f(\tsnr) |w|^2\right) \right\} \label{eq:asymp1}
\\
&= E_w \left\{ \lim_{m \to \infty} \log \left( 1 + f(\tsnr)
|w|^2\right) \right\} \label{eq:asymp2}
\\
&= E_w \left\{ \log \left( 1 + |w|^2 \lim_{m \to \infty} f(\tsnr)
\right) \right\} \label{eq:asymp3}
\\
&= E_w \left\{ \log \left( 1 + \tsnr |w|^2 \right) \right\}.
\label{eq:asymp4}
\end{align}
(\ref{eq:asymp1}) follows from the fact that $(m-1)/m \to 1$ as $m
\to \infty$. For (\ref{eq:asymp2}) to hold, we invoke the Dominated
Convergence Theorem \cite{Rudin}. Note that
\begin{align}
\left| \log(1 \!\!+\!\! f(\tsnr)|w|^2)\right| &\le f(\tsnr) |w|^2
\\
&= \frac{\phi(\tsnr)\tsnr^2}{\psi(\tsnr)\tsnr + (m-1)} |w|^2
\\
&\le \frac{\phi(\tsnr)}{\psi(\tsnr)}\tsnr |w|^2
\\
&= \frac{\delta^*(1-\delta^*)m^2}{m + m(m-2)\delta^*} \tsnr |w|^2
\\
&=\frac{(1-\delta^*)m^2}{\frac{m}{\delta^*} + m(m-2)} \tsnr |w|^2
\\
&\le \frac{(1-\delta^*)m}{m-2} \tsnr |w|^2 \label{eq:lastineq-1}
\\
&\le 3 \tsnr |w|^2 \quad \text{for } m\ge 3 \label{eq:lastineq}
\end{align}
where (\ref{eq:lastineq-1}) is obtained by removing
$\frac{m}{\delta^*}$ in the denominator and (\ref{eq:lastineq})
follows from the facts that $1-\delta^* \le 1$ and $\frac{m}{m-2}
\le 3$ for all $m \ge 3$. If $m=2$, we have $\phi(\tsnr) = 1$,
$\psi(\tsnr) = 2$, and hence
\begin{align}
\left| \log(1 \!\!+\!\! f(\tsnr)|w|^2)\right| &= \log\left(1 +
\frac{\tsnr^2}{2\tsnr+1}|w|^2 \right) \le \frac{1}{2}\tsnr |w|^2.
\end{align}
Therefore, $3 \tsnr |w|^2$ is an upper bound that applies for all
integer values $m \ge 2$. Furthermore, the upper bound does not
depend on $m$ and is integrable, i.e., $E_w\{3 \tsnr |w|^2\} =
3\tsnr <\infty$. Hence, the Dominated Convergence Theorem applies
and (\ref{eq:asymp2}) is justified. (\ref{eq:asymp3}) is due to the
fact that logarithm is a continuous function. (\ref{eq:asymp4}) can
easily be verified by noting that $m^2 \delta^*$ is the fastest
growing component, increasing as $m^{\frac{3}{2}}$ with increasing
$m$.

(\ref{eq:derivative}) follows again from the application of the
Dominated Convergence Theorem and the fact that the derivative of
$f(\tsnr)$ with respect to $\chi = 1/m$ at $\chi = 0$ is $-\infty$.
\hfill $\square$

\begin{figure}
\begin{center}
\includegraphics[width = \ffigsize\textwidth]{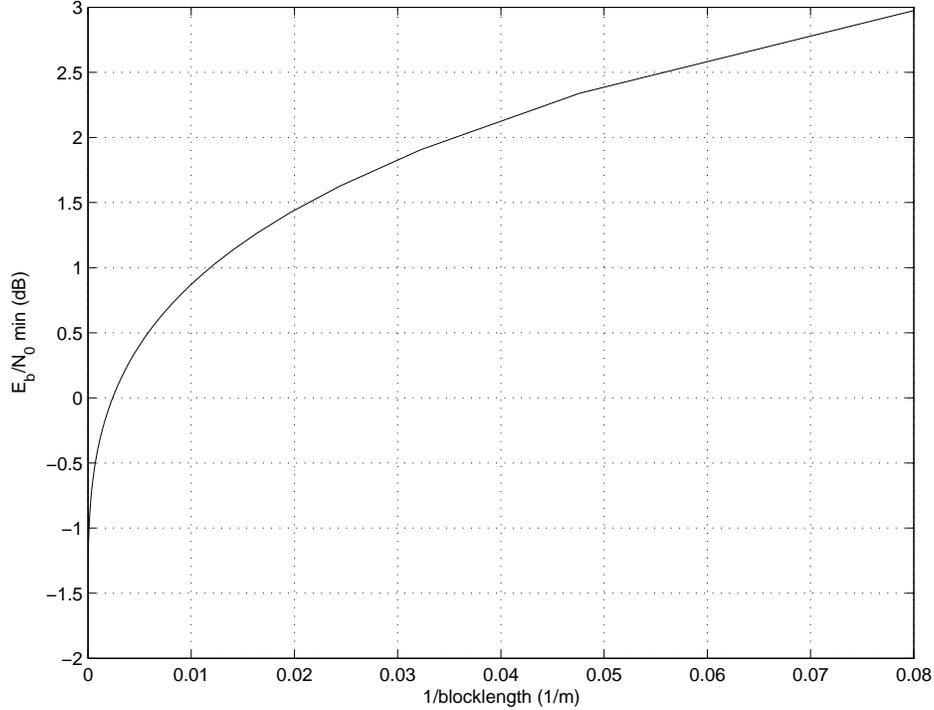}
\caption{Minimum energy per bit $\frac{E_{b,U}}{N_0}_{min}$ vs.
$\frac{1}{m}$ in the worst-case scenario} \label{fig:minbit_vs_m}
\end{center}
\end{figure}

The first part of Theorem \ref{prop:asympcap} is not surprising and
is expected because reference \cite{Marzetta} has already shown that
as the block length grows, the perfect knowledge capacity is
achieved even if no channel estimation is performed.
This result agrees with our observation in Fig.
\ref{fig:trainingbitenergy} that $-1.59$ dB is approached at lower
SNR values as $m$ increases. 
However, the rate of approach is very slow in terms of the block
size, as proven in the second part of Theorem \ref{prop:asympcap}
and evidenced in Fig. \ref{fig:minbit_vs_m}. Due to the infinite
slope\footnote{Note that Theorem \ref{prop:asympcap} implies that
the slope of $\frac{\tsnr}{C_L(\tsnr)}$ at $\chi = \frac{1}{m} = 0$
is $\infty$.} observed in the figure, approaching $-1.59$ dB is very
demanding in block length.

\subsection{Peak Power Constraint on the Pilot}

Heretofore, we have assumed that there are no peak power constraints
imposed on either the data or pilot symbols. Recall that the power
of the pilot symbol is given by
\begin{align}
|x_t|^2 &= \delta^*mP = \sqrt{\xi (\xi + mP)}-\xi
\label{eq:pilotpower}
\end{align}
where $\xi = \frac{m\gamma^2P + (m-1)N_0}{(m-2)\gamma^2}$. We
immediately observe from (\ref{eq:pilotpower}) that the pilot power
increases at least as $\sqrt{m}$ as $m$ increases. For large block
sizes, such an increase in the pilot power may be prohibitive in
practical systems. Therefore, it is of interest to impose a peak
power constraint on the pilot in the following form:
\begin{gather}
|x_t|^2 \le \kappa P.
\end{gather}
Since the average power is uniformly distributed over the data
symbols, the average power of a data symbol is proportional to $P$
and is at most $(1-\delta^*)2P$ for any block size. Therefore,
$\kappa$ can be seen as a limitation on the peak-to-average power
ratio. Note that we will allow Gaussian signaling for data
transmission. Hence, there are no hard peak power limitations on
data signals. This approach will enable us to work with a
closed-form capacity expression. Although Gaussian signals can
theoretically assume large values, the probability of such values is
decreasing exponentially. The case in which a peak power constraint
is imposed on both the training and data symbols is treated in the
Section \ref{sec:peakpower}.

If the optimal power allocated to a single pilot exceeds $\kappa P$,
i.e., $\delta^*mP > \kappa P \Rightarrow \delta^*m > \kappa$, the
peak power constraint on the pilot becomes active. In this case,
more than just a single pilot may be needed for optimal performance.

In this section, we address the optimization of the number of pilot
symbols when each pilot symbol has fixed power $|x_{t,i}|^2 =\kappa
P \,\, \forall i$. If the number of pilot symbols is $l < m$, then
$\|\x_t\|^2 = l\kappa P$ and, as we know from Section
\ref{sec:trainingscheme},
\begin{align}
\he \sim \CN \left( 0, \frac{\gamma^4 l\kappa P}{\gamma^2 l \kappa
P+ N_0}\right) \text{ and } \hn \sim \CN \left( 0, \frac{\gamma^2
N_0 }{\gamma^2 l \kappa P + N_0}\right). \nonumber
\end{align}
Similarly as before, when the estimate error is assumed to be
another source of additive noise and overall additive noise is
assumed to be Gaussian, the input-output mutual information achieved
by Gaussian signaling is given by
\begin{align}
I_{L,p} = \frac{m-l}{m}E_w\left\{ \log\left( 1 +
g(\tsnr,l)|w|^2\right)\right\}
\end{align}
where $w \sim \CN(0,1)$ and
\begin{gather}
g(\tsnr, l) = \frac{l\kappa(m - l \kappa) \tsnr^2}{(m - l\kappa +
(m-l)l\kappa)\tsnr + m-l}.
\end{gather}
\begin{figure}
\begin{center}
\includegraphics[width = \figsize\textwidth]{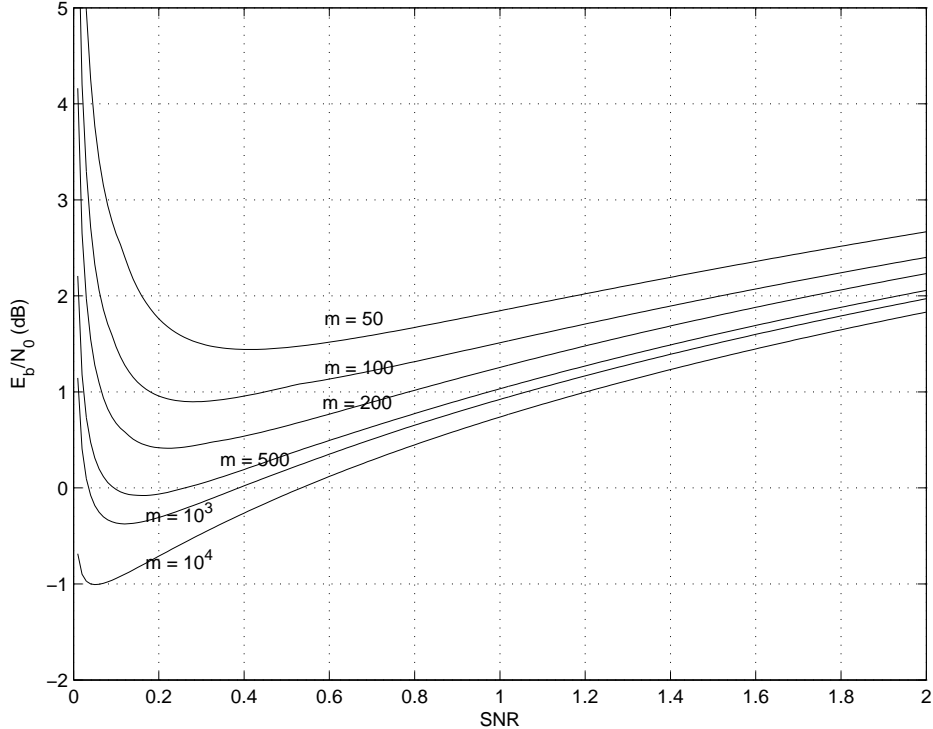}
\caption{Energy per bit $E_{b,U}/N_0$ vs. $\tsnr$ for block sizes of
$m = 50, 100, 200, 500, 10^3, 10^4$. The pilot peak power constraint
is $|x_t|^2 \le 10 P$.} \label{fig:ptrainingbitenergy}
\end{center}
\end{figure}
\begin{figure}
\begin{center}
\includegraphics[width = \figsize\textwidth]{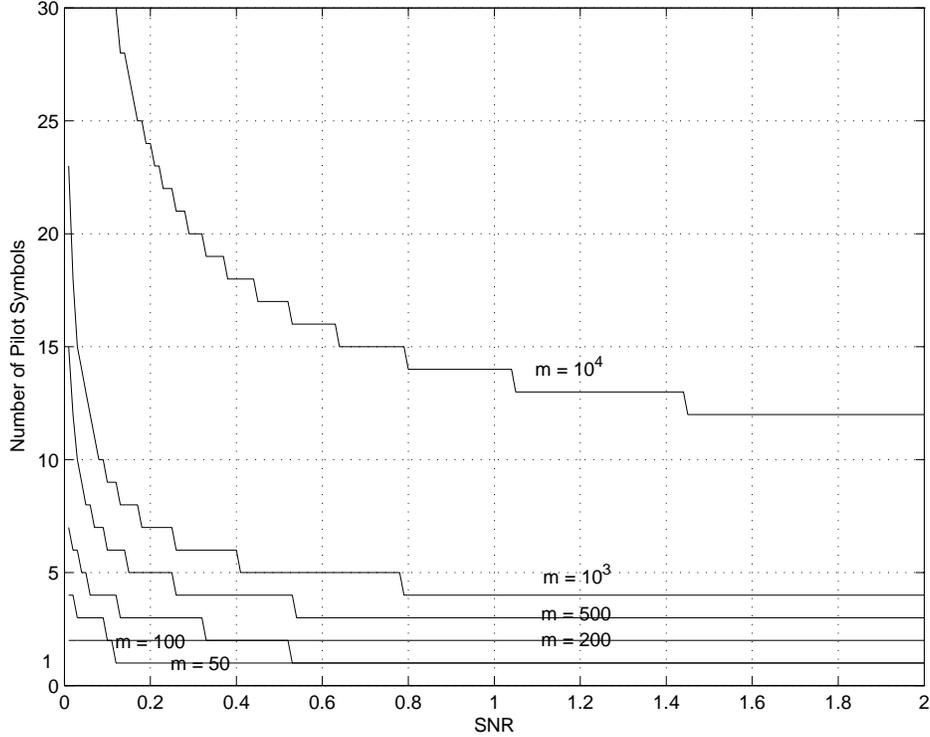}
\caption{Number of pilot symbols per block  vs. $\tsnr$ }
\label{fig:noofpilots}
\end{center}
\end{figure}
The optimal value of the training duration $l$ that maximizes
$I_{L,p}$ can be obtained through numerical optimization. Fig.
\ref{fig:ptrainingbitenergy} plots the normalized bit energy values
$\frac{\tsnr \log2}{I_{L,p}}$ in dB obtained with optimal training
duration for different block lengths. The peak power constraint
imposed on a pilot symbol is $|x_{t}|^2 \le 10 P$. Fig.
\ref{fig:noofpilots} gives the optimal number of pilot symbols per
block. From Fig. \ref{fig:ptrainingbitenergy}, we observe that the
minimum bit energy, which is again achieved at a nonzero value of
the $\tsnr$, decreases with increasing block length and approaches
to the fundamental limit of $-1.59$ dB. We note from Fig.
\ref{fig:noofpilots} that the number pilot symbols per block
increases as the block length increases or as $\tsnr$ decreases to
zero.
When there are no peak constraints, $\delta^* \to 1/2$ as $\tsnr \to
0$. Hence, we need to allocate approximately half of the available
total power $mP$ to the single pilot signal in the low-power regime,
increasing the peak-to-average power ratio. In the limited peak
power case, this requirement is translated to the requirement of
more pilot symbols per block at low $\tsnr$ values.

\begin{table}
\caption{} \label{table:minbitenergy}
\begin{center}\begin{tabular}{|c|c|c|c|c|} \hline
&  $\frac{E_{b,U}}{N_0}_\text{min}$(dB) & \# of pilots & $\tsnr$ & $\frac{E_{b,U}}{N_0}_\text{min}$ (dB) (\scriptsize{no peak constraints})\\
\hline $m = 50$ & 1.441 & 1 & 0. 41& 1.440
\\ \hline
$m = 100$ & 0.897 & 2 & 0.28& 0.871
\\ \hline
$m = 200$ & 0.413 & 3 & 0.22& 0.404
\\ \hline
$m = 500$ & -0.079 & 5 & 0.16& - 0.085
\\ \hline
$m = 10^3$ & -0.375 & 9 & 0.12& -0.378
\\ \hline
$m = 10^4$ & -1.007 & 44 & 0.05 & -1.008
\\ \hline
\end{tabular}
\end{center}
\end{table}

Table \ref{table:minbitenergy} lists, for different values of $m$,
the minimum bit energy values, the required number of pilot symbols
at this level, and the $\tsnr$ at which minimum bit energy is
achieved. It is again assumed that $\kappa = 10$. The last column of
the table provides the minimum bit energy attained when there are no
peak power constraints on the pilot signal. As the block size
increases, the minimum bit energy is achieved at a lower $\tsnr$
value while a longer training duration is required. Furthermore,
comparison with the last column indicates that the loss in minimum
bit energy incurred by the presence of peak power constraints is
negligible. The following result shows that the capacity of the
perfectly known channel, and hence the minimum bit energy of
$-1.59$dB, is approached with simultaneous growth of training
duration and block length. Note that this result conforms with the
results in Table \ref{table:minbitenergy}.

\begin{prop:pasympcap}

Assume that the training duration $l(m,\tsnr)$ increases as $m$
increases and satisfies
\begin{gather}
\lim_{m \to \infty} \frac{l(m,\tsnr)}{m} = 0.
\label{eq:condinpasympcap}
\end{gather}
Then, $ \quad \lim_{m \to \infty} I_{L,p} = E_{w}\{\log(1 + \tsnr
|w|^2)\}. $
\end{prop:pasympcap}

\emph{Proof}: We have
\begin{align}
\lim_{m \to \infty} I_{L,p} &= \lim_{m \to \infty} \left(1
-\frac{l}{m}\right)E_w\left\{ \log\left( 1 +
g(\tsnr,l)|w|^2\right)\right\} \nonumber
\\
&= \lim_{m \to \infty} E_w\left\{ \log\left( 1 +
g(\tsnr,l)|w|^2\right)\right\} \label{eq:pasymp1}
\\
&= E_w\left\{ \lim_{m \to \infty }\log\left( 1 +
g(\tsnr,l)|w|^2\right)\right\} \label{eq:pasymp2}
\\
&=E_w\{\log(1+\tsnr |w|^2)\}. \label{eq:pasymp3}
\end{align}
(\ref{eq:pasymp1}) follows from the condition
(\ref{eq:condinpasympcap}). (\ref{eq:pasymp2}) can be justified by
invoking the Dominated Convergence Theorem \cite{Rudin} similarly as
in the proof of Theorem \ref{prop:asympcap}. (\ref{eq:pasymp3})
follows from
\begin{gather}
\lim_{m \to \infty} g(\tsnr, l) = \tsnr,
\end{gather}
which holds if the conditions of the theorem are met. \hfill
$\square$

\subsection{Flash Training and Transmission}
\label{subsec:worstcaseflash}

One approach to improve the energy efficiency in the low $\tsnr$
regime is to increase the peak power of the transmitted signals.
This can be achieved by transmitting $\nu$ fraction of the time with
power $P/\nu$. Note that training also needs to be performed only
$\nu$ fraction of the time. In this section, no peak power
constraints are imposed on pilot symbols. This type of training and
communication, called flash training and transmission scheme, is
analyzed in \cite{Zheng2} where it is shown that the minimum bit
energy of $-1.59$ dB can be achieved if the block length $m$
increases at a certain rate as $\tsnr$ decreases. In the setting we
consider, flash transmission scheme achieves the following rate:
\begin{gather}
C_{fL}(\tsnr, \nu) = \nu(\tsnr) C_L\left(
\frac{\tsnr}{\nu(\tsnr)}\right)
\end{gather}
where $0 < \nu(\cdot) \le 1$ is the duty cycle which in general is a
function of the $\tsnr$. First, we show that flash transmission
using peaky Gaussian signals does not improve the minimum bit
energy.
\begin{prop:flashminbitenergy} \label{prop:flashminbitenergy}
For any duty cycle function $\nu(\cdot)$,
\begin{gather}
\inf_\tsnr \frac{\tsnr}{C_{fL}(\tsnr,\nu)} \ge \inf_\tsnr
\frac{\tsnr}{C_{L}(\tsnr)}.
\end{gather}
\end{prop:flashminbitenergy}

\vspace{.5cm} \emph{Proof}: Note that for any $\tsnr$ and
$\nu(\tsnr)$,
\begin{align}
\!\!\!\!\!\!\!\!\!\!\!\!\!\!\frac{\tsnr}{C_{fL}(\tsnr,\nu)} =
\frac{\frac{\tsnr}{\nu(\tsnr)}}{C_L\left(
\frac{\tsnr}{\nu(\tsnr)}\right)}
=\frac{\tilde{\tsnr}}{C_L(\tilde{\tsnr})} \ge \inf_\tsnr
\frac{\tsnr}{C_{L}(\tsnr)} \label{eq:flashineq}
\end{align}
where $\tilde{\tsnr}$ is defined as the new $\tsnr$ level. Since the
inequality in (\ref{eq:flashineq}) holds for any $\tsnr$ and
$\nu(\cdot)$, it also holds for the infimum of the left-hand side of
(\ref{eq:flashineq}), and hence the result follows. \hfill $\square$

We classify the duty cycle function into three categories:
\begin{enumerate}
\item $\nu(\cdot)$ that satisfies $\lim_{\tsnr \to 0} \frac{\tsnr}{\nu(\tsnr)} = 0$
\item $\nu(\cdot)$ that satisfies $\lim_{\tsnr \to 0} \frac{\tsnr}{\nu(\tsnr)} = \infty$
\item $\nu(\cdot)$ that satisfies $\lim_{\tsnr \to 0}
\frac{\tsnr}{\nu(\tsnr)} = a$ for some constant $a > 0$.
\end{enumerate}
Next, we analyze the performance of each category of duty cycle
functions in the low-$\tsnr$ regime.

\begin{prop:flashbitenergy} \label{prop:flashbitenergy}
If $\nu(\cdot)$ is chosen from either Category 1 or 2,
\begin{gather}
\left.\frac{E_{b,U}}{N_0}\right|_{C_{fL} = 0} = \lim_{\tsnr \to 0}
\frac{\tsnr}{C_{fL}(\tsnr, \nu)}\log2 = \infty.
\end{gather}
If $\nu(\cdot)$ is chosen from Category 3,
\begin{align}
\left.\frac{E_{b,U}}{N_0}\right|_{C_{fL} = 0} &= \frac{m}{m-1}
\frac{a}{E_w\{\log_2(1 + f(a)|w|^2)\}}.
\end{align}
\end{prop:flashbitenergy}

\emph{Proof}: We first note that by Jensen's inequality,
\begin{align}
\frac{C_{fL}(\tsnr, \nu)}{\tsnr} &\le \frac{m\!-\!\!1}{m}
\frac{\nu(\tsnr)}{\tsnr}\log \left( 1 + f\left(
\frac{\tsnr}{\nu(\tsnr)}\right)\right)
\label{eq:cappercostupperbound}
\\
&\stackrel{\text{def}}{=} \zeta(\tsnr, \nu).
\label{eq:cappercostupperbounddef}
\end{align}
First, we consider category 1. In this case, as $\tsnr \to 0$,
$\frac{\tsnr}{\nu(\tsnr)} \to 0$. As shown before, the logarithm in
(\ref{eq:cappercostupperbound}) scales as
$\frac{\tsnr^2}{\nu(\tsnr^2)}$ as $\tsnr \to 0$, and hence
$\zeta(\tsnr,\nu)$ scales as $\frac{\tsnr}{\nu(\tsnr)}$ leading to
\begin{gather}
\lim_{\tsnr \to 0} \frac{C_{fL}(\tsnr,\nu)}{\tsnr} \le \lim_{\tsnr
\to 0} \zeta(\tsnr, \nu) = 0.
\end{gather}
In category 2, $\frac{\tsnr}{\nu(\tsnr)}$ grows to infinity as
$\tsnr \to 0$. Since the $\log(\cdot)$ function on the right hand
side of (\ref{eq:cappercostupperbound}) increases only
logarithmically as $\frac{\tsnr}{\nu(\tsnr)} \to \infty$, we can
easily verify that
\begin{gather}
\lim_{\tsnr \to 0} \frac{C_{fL}(\tsnr,\nu)}{\tsnr} \le \lim_{\tsnr
\to 0} \zeta(\tsnr, \nu) = 0.
\end{gather}
In category 3, $\nu(\tsnr)$ decreases at the same rate as $\tsnr$.
In this case, we have
\begin{align}
\lim_{\tsnr \to 0} \!\!\!\!\frac{C_{fL}(\tsnr,\nu)}{\tsnr} &=
\lim_{n \to \infty}
\frac{C_{fL}\left(\frac{1}{n},\nu\right)}{\frac{1}{n}}
\\
&=\frac{\frac{m-1}{m}E_w\{\lim_{n \to \infty} \log\left(1 +
f(\frac{1}{nv})|w|^2 \right)\}}{a} \label{eq:categ12}
\\
&=\frac{\frac{m-1}{m}E_w\{\log\left(1 + f(a)|w|^2 \right)\}}{a}
\label{eq:categ13}
\end{align}
(\ref{eq:categ12}) is justified by invoking the Dominated
Convergence Theorem and noting the integrable upper bound
\begin{align}
\left|\log\left(1+f\left(\frac{1}{n\nu}\right)|w|^2\right)\right|
\le 3 \frac{1}{n\nu} |w|^2 \le \frac{3}{\nu} |w|^2 \text{ for } n
\ge 1. \nonumber
\end{align}
The above upper bound is given in the proof of Theorem
\ref{prop:asympcap}. Finally, (\ref{eq:categ13}) follows from the
continuity of the logarithm. \hfill $\square$

Theorem \ref{prop:flashbitenergy} shows that if the rate of the
decrease of the duty cycle  is faster or slower than $\tsnr$ as
$\tsnr \to 0$, the bit energy requirement still increases without
bound in the low-$\tsnr$ regime. This observation is tightly linked
to the fact that the capacity curve $C_L$ has a zero slope as both
$\tsnr \to 0$ and $\tsnr \to \infty$. For improved performance in
the low-$\tsnr$ regime, it is required that the duty cycle scale as
$\tsnr$. A particularly good choice is
$$\nu(\tsnr) = \frac{1}{a^*} \tsnr$$ where $a^*$ is
equal to the $\tsnr$ level at which the minimum bit energy is
achieved in a non-flashy transmission scheme. With this choice, we
basically perform time-sharing between $\tsnr = 0$ and $\tsnr =
a^*$.
\begin{figure}
\begin{center}
\includegraphics[width = \ffigsize\textwidth]{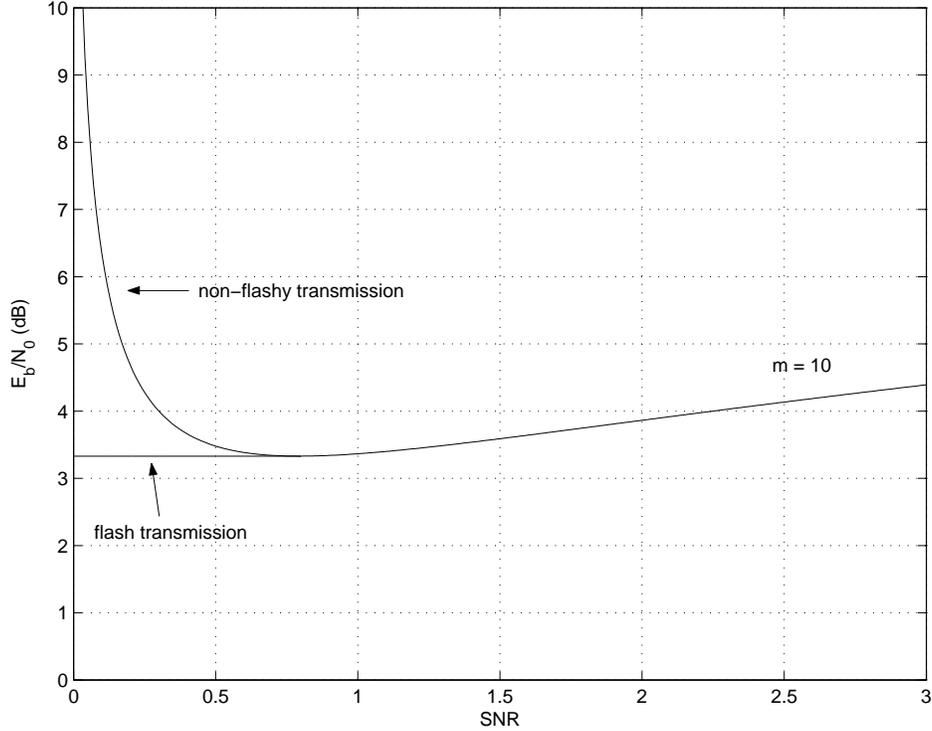}
\caption{Energy per bit $E_{b,U}/N_0$ vs. $\tsnr$ for non-flashy and
flash transmissions.} \label{fig:flashbitenergy}
\end{center}
\end{figure}
Fig. \ref{fig:flashbitenergy} plots the normalized bit energy
$\frac{E_{b,U}}{N_0}$ as a function of $\tsnr$ for block size $m =
10$. The minimum bit energy is achieved at $\tsnr= 0.8$. For $\tsnr
< 0.8$, flash transmission is employed with $\nu(\tsnr) = 1/0.8 \,
\tsnr$. As observed in the figure, the minimum bit energy level can
be maintained for lower values of $\tsnr$ at the cost of increased
peak-to-average power ratio. It should be noted that the optimal
point of operation is still at $\tsnr = 0.8$ since operating at
$\tsnr < 0.8$ will result in reduced data rates without any
improvements in the bit energy. From a different perspective, if
$\tsnr$ is the signal-to-noise ratio per unit bandwidth, then
increasing the bandwidth so that $\tsnr < 0.8$ will not produce any
energy savings. However, in circumstances in which regulations or
device properties dictate operation at $\tsnr$ values lower than the
minimum bit energy point, flash transmission can be adopted to
improve the energy efficiency.

\section{Capacity and Energy Efficiency in the Presence of Peak Power
Limitations} \label{sec:peakpower}

In this section, we consider the channel
\begin{gather} \label{eq:dataphasemodel2}
\y_d = \he \x_d + \hn \x_d + \n_d
\end{gather}
and assume that the channel input is subject to the following peak
power constraint
\begin{gather} \label{eq:peakpower}
\|\x\|^2 \stackrel{\text{a.s.}}{\le} mP.
\end{gather}
In this setting, it is again easy to see that the transmission of a
single pilot is optimal. Since the peak power constraint is imposed
on the input vector $\x$, the pilot power can be varied instead of
increasing the number of pilot symbols. Similarly as before, we
assume that the pilot symbol power is
\begin{gather}
|x_t|^2 = \delta mP.
\end{gather}
Therefore, the ($m-1$)-dimensional data vector $\x_d$ is subject to
\begin{gather} \label{eq:datapeakpower}
\|\x_d\|^2 \stackrel{\text{a.s.}}{\le} (1-\delta)mP.
\end{gather}
Our goal is to solve the maximization problem
\begin{gather} \label{eq:pcap}
C = \sup_{\delta \in (0,1)} \sup_{\substack{\x_d \\ \|\x_d\|^2
\stackrel{\text{a.s.}}{\le} (1-\delta)mP} } \frac{1}{m}\,I(\x_d;
\y_d | \he)
\end{gather}
and obtain the channel capacity, and identify the capacity-achieving
input distribution and the optimal value of the power allocation
coefficient $\delta$. The input-output mutual information is
\begin{gather} \label{eq:mutualinfo}
I(\x_d ; \y_d | \he) = E_{\he} E_{\x_d} \int f_{\y|\x_d, \he}(\y |
\x_d, \he) \log \frac{f_{\y|\x_d, \he}(\y | \x_d, \he)}{f_{\y|
\he}(\y |\he)} \, d \y
\end{gather}
where
\begin{gather} \label{eq:condpdf}
f_{\y|\x_d, \he}(\y | \x_d, \he) = \frac{\exp\left(-(\y - \he\x_d
)^{\dagger}(\tgamma^2 \x_d \x_d^{\dagger} + N_0 \I)^{-1} (\y
-\he\x_d )\right)}{\pi^{m-1} N_0^{m-2}(\tgamma^2 \|\x_d\|^2 + N_0)}
\end{gather}
and
\begin{gather}
\tgamma^2 = E\{|\hn|^2\} = \frac{\gamma^2 N_0}{\gamma^2 \delta mP +
N_0}.
\end{gather}
First, we have the following preliminary result on the structure of
the capacity-achieving input distribution.
\begin{prop:blisotropic} \label{prop:blisotropic}
For the block fading channel (\ref{eq:dataphasemodel2}) where the
input is subject to a peak power limitation
(\ref{eq:datapeakpower}), the capacity-achieving input vector can be
written as $\x_d = \|\x_d\| \bv$ where $\|\x_d\|$ is a nonnegative
real random variable and $\bv$ is an independent isotropically
distributed unit random vector.
\end{prop:blisotropic}
\emph{Proof}: The proof follows primarily from the same techniques
developed in \cite{Marzetta}. First note the invariance of the peak
constraint (\ref{eq:datapeakpower}) to rotations of the input. Since
$f_{\y|\x}(\Phi \y|\Phi \x_d, \he) = f_{\y|\x_d, \he}(\y | \x_d,
\he)$ for any $(m-1) \times (m-1)$ dimensional deterministic unitary
matrix $\Phi$, it can be easily seen that the mutual information is
also invariant to deterministic rotations of the input, and the
result follows from the concavity of the mutual information which
implies that there is no loss in optimality if one uses circularly
symmetric input distributions. \hfill $\square$

With this characterization, the problem has been reduced to the
optimization of the input magnitude distribution, $F_{\x_d}$. We
first obtain an equivalent expression for the mutual information
when the the input vector has the structure described in Theorem
\ref{prop:blisotropic}.

\begin{prop:eqvmutual} \label{prop:eqvmutual}
When the input is $\x_d = \|\x_d\| \bv$ where $\bv$ is an
isotropically distributed unit vector that is independent of the
magnitude $\|\x_d\|$, the input-output mutual information of the
channel (\ref{eq:dataphasemodel2}) can be expressed as
\begin{align} \label{eq:mutualinfoisotrop}
I(\x_d; \y_d | \he ) = I(F_r | \he) = -E_{\K, r} \left\{
\int_0^\infty f_{R|r,\K}(R|r,\K) \log \,g(R,F_r, \K ) \, dR \right\}
- E_r \{\log(1+r^2)\} - (m-1)
\end{align}
where
\begin{gather}
f_{R|r,\K}(R|r,\K) = \left\{
\begin{array}{ll}
\frac{R^{m-2}}{(m-3)!} \, \frac{e^{-R - \frac{\K
r^2}{1+r^2}}}{1+r^2} \int_0^1 (1-a)^{m-3} e^{\frac{ar^2R}{1+r^2}} \,
I_0\left(\frac{2\sqrt{\K R}\,r\sqrt{a}}{1+r^2} \right) \, da & m\ge
3
\\
\frac{e^{- \frac{R + \K r^2}{1+r^2}}}{1+r^2}
I_0\left(\frac{2\sqrt{\K R}\,r}{1+r^2} \right)& m = 2
\end{array} \right. \label{eq:fRr}
\\ \intertext{and}
g(R,F_r, \K) = \frac{(m-2)!}{R^{m-2}} \int_0^\infty
f_{R|r,\K}(R|r,\K) \, dF_r. \label{eq:gR}
\end{gather}
In the above formulations, $R = \frac{\|\y\|^2}{N_0}$, $r =
\frac{\tgamma \|\x_d\|}{\sqrt{N_0}}$, and $\K =
\frac{|\he|^2}{\tgamma^2}$. Furthermore, $F_r$ denotes the
distribution function of $r$. $\K$ is an exponential random variable
with mean $E\{\K\} = \frac{E\{|\he^2|\}}{\tgamma^2} = \frac{\gamma^2
\delta m P}{N_0}$. $E_{\K,r}$ denotes the expectation with respect
to $\K$ and $r$.
\end{prop:eqvmutual}

\emph{Proof}: See Appendix \ref{app:eqvmutual}.

Note that the integral in the mutual information expression in
(\ref{eq:mutualinfo}) is in general an $2(m-1)$-fold integral. In
(\ref{eq:mutualinfoisotrop}), this has been reduced to a double
integral providing a significant simplification especially for
numerical analysis. With this result, the channel capacity in nats
per symbol can now be reformulated as
\begin{gather} \label{eq:pcapr}
C = \sup_{\delta \in (0,1)} C_\delta  = \sup_{\delta \in (0,1)}
 \sup_{\substack{F_r \\ r \stackrel{\text{a.s.}}{\le} \sqrt{L}} }
\frac{1}{m} \, I(F_r | \he)
\end{gather}
where $L = \frac{\gamma^2(1-\delta)mP}{\gamma^2 \delta mP + N_0}$.
Hence, the capacity is obtained through the optimal choices of the
power allocation coefficient $\delta$ and normalized input magnitude
distribution $F_r$. Since the inner maximization is over a
continuous alphabet, the existence of the capacity-achieving
distribution $F_r$ is not guaranteed. Next, we prove the existence
of a capacity-achieving input distribution and provide a sufficient
and necessary condition for an input to be optimal.

\begin{prop:existence+ktc} \label{prop:existence+ktc}
Fix the value of $\delta \in (0,1)$ and consider the inner
maximization in (\ref{eq:pcapr}). There exists an input distribution
$F_r$ that maximizes the mutual information $I(F_r|\he)$. Moreover,
an input distribution $F_r$ is capacity-achieving if and only if the
following Kuhn-Tucker condition is satisfied:
\begin{gather} \label{eq:ktc}
\Phi(r) = E_{\K} \left\{ \int_0^\infty f_{R|r,\K}(R|r,\K) \log
\,g(R,F_r, \K ) \, dR \right\} + \log(1+r^2) + mC_\delta + (m-1) \ge
0 \quad \forall r \in [0,\sqrt{L}]
\end{gather}
with equality at the points of increase of $F_r$\footnote{The set of
points of increase of a distribution function $F$ is $\{r:
F(r-\epsilon) < F(r + \epsilon) \,\,\, \forall \epsilon >0$\}.}. In
the above condition, $C_\delta$ denotes the result of the inner
maximization in (\ref{eq:pcapr}) .
\end{prop:existence+ktc}

\emph{Proof}: See Appendix \ref{app:existence+ktc}.

Having shown the existence of the capacity-achieving input
distribution and a sufficient and necessary condition for an input
distribution to be optimal, we turn our attention to the
characterization of the optimal input.

\begin{theo:discrete} \label{theo:discrete}
Fix the value of $\delta \in (0,1)$. The input distribution that
maximizes the mutual information $I(F_r|\he)$ is discrete with a
finite number of mass points
\end{theo:discrete}

\emph{Proof}: The following upper bound is obtained in Appendix
\ref{app:existence+ktc}:
\begin{align}
g(R,F, \K ) &\le (m-2)\, e^{-\frac{R}{1+L} + \sqrt{\K R}}
\label{eq:uppg3sect}
\end{align}
Using this upper bound, we have
\begin{align}
\!\!\!\!E_{\K} \left\{ \int_0^\infty \!\!\!\!\!f_{R|r,\K}(R|r,\K)
\log \,g(R,F_r, \K ) \, dR \right\} &= E_\K E_{R|r,\K} \{\log
\,g(R,F_r, \K )\} \label{eq:uppktcinteg1}
\\
&\le \log(m\!-\!2) -E_\K E_{R|r,\K}\left\{\frac{R}{1+L}\right\} +
E_\K E_{R|r,\K}\left\{\sqrt{\K R}\right\} \label{eq:uppktcinteg2}
\\
&\le \log(m\!-\!2) -E_\K E_{R|r,\K}\left\{\frac{R}{1+L}\right\} +
E_\K \left\{\sqrt{\K} \sqrt{E_{R|r,\K}\{R\}}\right\}
\label{eq:uppktcinteg3}
\\
&\le \log(m\!-\!2) -\frac{(1+E_\K\{\K\})r^2+m-1}{1+L} \nonumber
\\
&\,\,\,\,\,\,\,+ E_\K \left\{\sqrt{\K} \sqrt{(1+\K)r^2+m-1}\right\}.
\label{eq:uppktcinteg4}
\end{align}
(\ref{eq:uppktcinteg2}) follows from (\ref{eq:uppg3sect}), and
(\ref{eq:uppktcinteg3}) follows from the fact that $E\{\sqrt{R}\}
\le \sqrt{E\{R\}}$. Finally, (\ref{eq:uppktcinteg4}) is obtained by
noting that $E_{R|r,\K}\{R\} = (1+\K)r^2 + m-1$. Note that the upper
bound in (\ref{eq:uppktcinteg4}), and hence the left-hand-side of
(\ref{eq:ktc}), decreases to $-\infty$ as $r \to \infty$ due to the
presence of $-r^2$ in the second term.

We prove the result by contradiction. Hence, we now assume that the
optimal input distribution $F_0$ has an infinite number of points of
increase on a bounded interval. Next, we extend the $\Phi(\cdot)$ in
(\ref{eq:ktc}) to the complex domain:
\begin{gather} \label{eq:ktccomplex}
\Phi(z) = E_{\K} \left\{ \int_0^\infty f_{R|r,\K}(R|z,\K) \log
\,g(R,F_r, \K ) \, dR \right\} + \log(1+z^2) + C + (m-1)
\end{gather}
where $z \in \mathbb{C}$ and $\log$ is the principle branch of the
logarithm. The Identity Theorem for analytic functions \cite{Knopp}
states that if two functions are analytic in a region and if they
coincide for an infinite number of distinct points having a limiting
point, they are equal everywhere in that region. It is shown in
Appendix \ref{app:analyticity} that $\Phi(z)$ is analytic in a
region $\mathcal{D}$ that includes the positive real line. By the
above assumption on the optimal input distribution, $\Phi(z) = 0$
for an infinite number of points having a limiting
point\footnote{The Bolzano-Weierstrass Theorem \cite{Rudin} states
that every bounded infinite set of real numbers has a limit point.}
in region $\mathcal{D}$. Therefore, by the Identity Theorem, we
should have $\Phi(r) = 0$ for all $r \ge 0$.
Clearly, this is not possible from the upper bound in
(\ref{eq:uppktcinteg3}) which diverges to $-\infty$ as $r \to
\infty$. Hence, the optimal input cannot have an infinite number of
points of increase on a bounded interval, from which we conclude
that the optimal input distribution is discrete with a finite number
of mass points. \hfill $\square$

\begin{figure}
\begin{center}
\includegraphics[width = \figsize\textwidth]{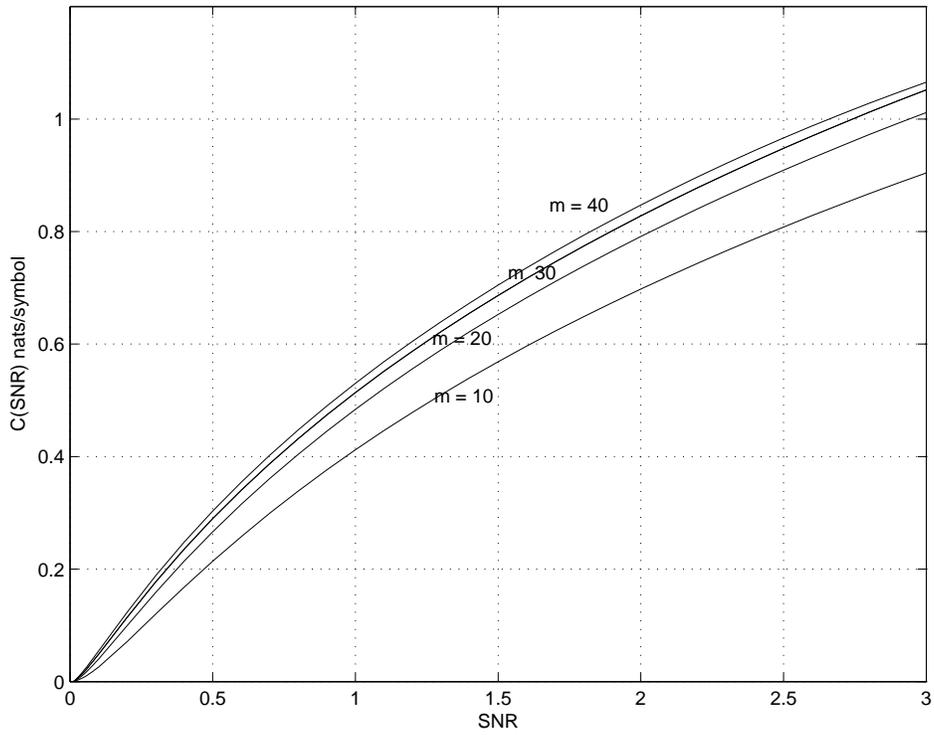}
\caption{Capacity (nats/symbol) vs. $\tsnr$ for block lengths of $m
= 10,20, 30$ and $40$ when the input is subject to peak power
limitations.} \label{fig:cap}
\end{center}
\end{figure}

\begin{figure}
\begin{center}
\includegraphics[width = \figsize\textwidth]{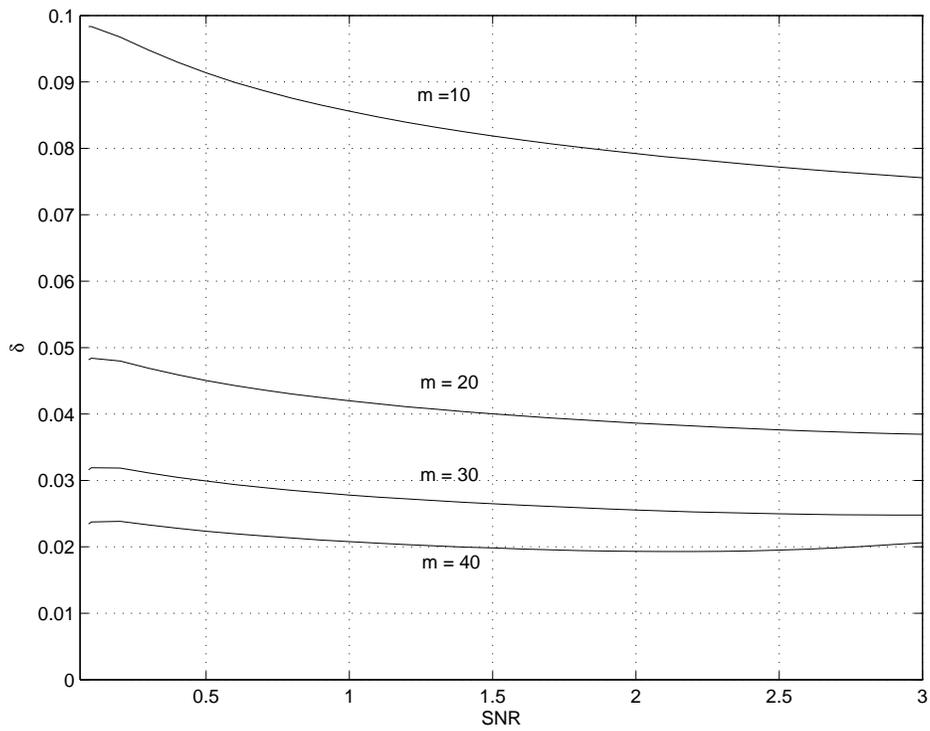}
\caption{Optimal fraction of power $\delta$ allocated to the pilot
symbol vs. $\tsnr$ for block lengths of $m = 10,20,30$ and $40$.}
\label{fig:delta}
\end{center}
\end{figure}

After the characterization of the discrete nature of the optimal
input, the optimization problem in (\ref{eq:pcapr}) can be solved
using vector optimization techniques. Numerical results indicate
that the optimal magnitude distribution $F_r$ has a single mass at
the peak level $r = \sqrt{L}$ for low-to-medium received peak $\tsnr
= \frac{\gamma^2 P}{N_0}$ levels. Hence, all the information is
carried by the isotropically distributed directional unit vector.
Therefore, information transmission is achieved by sending points on
the surface of an $(m-1)$-dimensional complex sphere with radius
$\frac{\sqrt{L N_0}}{\tgamma}$. Note that the mutual information (in
nats per $m$ symbols) achieved by having a single-mass at $r =
\sqrt{L}$ is
\begin{gather}\label{eq:pcaprsingle}
I_{cm} = -E_{\K} \left\{ \int_0^\infty f_{R|r,\K}(R|r=\sqrt{L},\K)
\log \,g(R,F_r, \K ) \, dR \right\} - \log(1+L) - (m-1).
\end{gather}
Figure \ref{fig:cap} plots the capacity values as a function of
$\tsnr$ for block lengths of $m = 10,20,30$ and $40$. These capacity
values are achieved with optimal power allocation. The optimal
fractions of power allocated to the pilot symbol are plotted in Fig.
\ref{fig:delta}. Note that for the range of $\tsnr$ values
considered in the figure, the optimal value of $\delta$ is slightly
smaller than $1/m$ and approaches $1/m$ as $\tsnr$ tends to 0. This
power allocation strategy is significantly different from that of
the worst-case scenario in which $\delta^* \to 1/2$ with decreasing
$\tsnr$.

\begin{figure}
\begin{center}
\includegraphics[width = \figsize\textwidth]{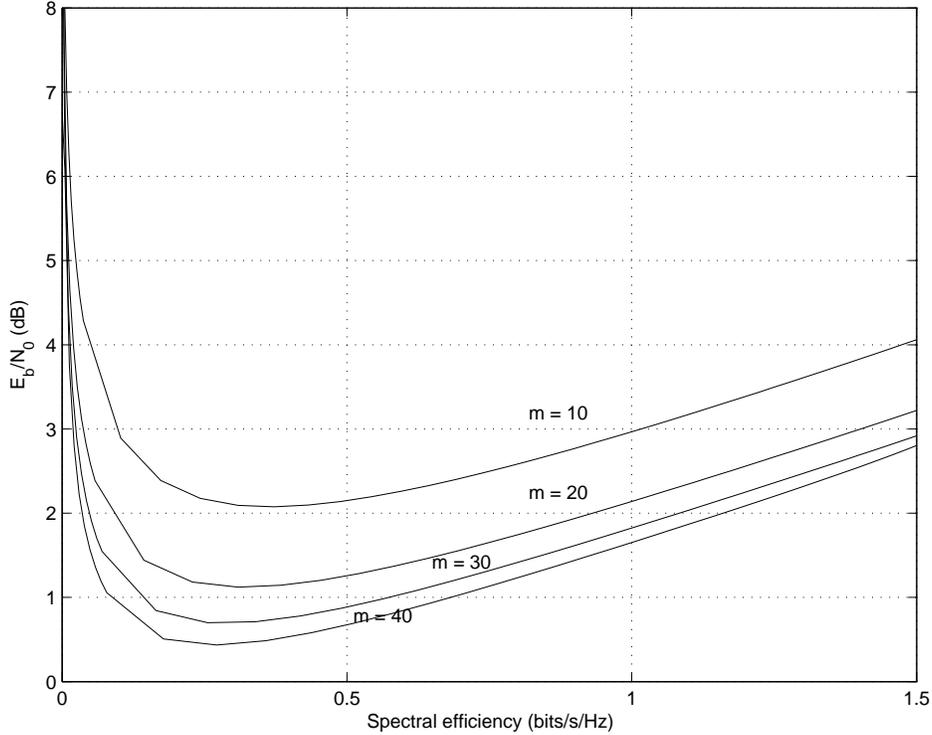}
\caption{Bit energy $\frac{E_b}{N_0}$ vs. Spectral efficiency
$\C\left(\frac{E_b}{N_0}\right)$ in pilot-assisted systems with
block lengths $m = 10,20,30$ and $40$.} \label{fig:ebno}
\end{center}
\end{figure}

In the low-$\tsnr$ regime, the tradeoff between spectral efficiency
and energy per bit obtained from $\frac{E_b}{N_0} = \frac{\tsnr
\log2}{C(\tsnr)}$ is the key performance measure \cite{Verdu}. If we
assume, without loss of generality, that one symbol occupies a $1
\text{s} \times 1 \text{Hz}$ time-frequency slot, then the maximum
spectral efficiency is $\C(E_b/N_0) = C(\tsnr)\log_2e$ bits/s/Hz
where we have assumed that $C(\tsnr)$ is in nats/symbol. Fig.
\ref{fig:ebno} plots the bit energy values as a function of the
spectral efficiency. It is again observed that the minimum bit
energy is achieved at a nonzero spectral efficiency and the required
bit energy values grow without bound as $\tsnr$ and hence the
spectral efficiency is further decreased. Indeed, we can show the
following result.
\begin{theo:bitenergysinglemass} \label{theo:bitenergysinglemass}
Assume that the normalized input magnitude distribution has a single
mass and hence the magnitude is fixed at $r = \sqrt{L}$. For any
value of $\delta \in (0,1)$, the normalized bit energy required by
this input grows without bound as the signal-to-noise ratio
decreases to zero, i.e.,
\begin{gather}
\left.\frac{E_{b,cm}}{N_0}\right|_{I_{cm} = 0} = \lim_{\tsnr \to 0}
\frac{m\tsnr}{I_{cm}(\tsnr)}\log2 = \frac{m\log2}{\dot{I}_{cm}(0)} =
\infty.
\end{gather}
\end{theo:bitenergysinglemass}
\vspace{0.2cm} \emph{Proof:} Recall that $L =
\frac{(1-\delta)m\tsnr}{\delta m \tsnr + 1}$ and $\tsnr =
\frac{\gamma^2 P}{N_0}$. Also, note that an expression for $I_{cm}$
is given in (\ref{eq:pcaprsingle}). By making a change of variables,
we have the following equivalent expression:
\begin{gather}
I_{cm} = -E_{\K} \left\{ \int_0^\infty
f_{R|r,\K}(R|\sqrt{L},\K\delta m \tsnr) \log \,g(R,F_r, \K\delta m
\tsnr ) \, dR \right\} - \log(1+L) - (m-1)
\end{gather}
where $\K$ is now an exponential random variable with $E\{\K\} = 1$,
and hence is independent of $\tsnr$. We can easily show that
\begin{align}
\left.\frac{\partial}{\partial \tsnr}f_{R|r,\K}(R|\sqrt{L},\K\delta
m \tsnr)\right|_{\tsnr = 0} = -(1-\delta)m \frac{R^{m-2}}{(m-2)!} \,
e^{-R}  + (1-\delta)m \frac{R^{m-1}}{(m-1)!} \, e^{-R}.
\end{align}
Note that $$g(R,F_r, \K\delta m \tsnr) = \frac{(m-2)!}{R^{m-2}}
f_{R|r,\K}(R|\sqrt{L},\K\delta m \tsnr).$$ Using these facts, we can
easily prove that
\begin{gather}
\dot{I}_{cm}(0) = \left.\frac{\partial I_{cm}}{\partial
\tsnr}\right|_{\tsnr = 0} = 0.
\end{gather}
\hfill $\square$

\begin{figure}
\begin{center}
\includegraphics[width = \figsize\textwidth]{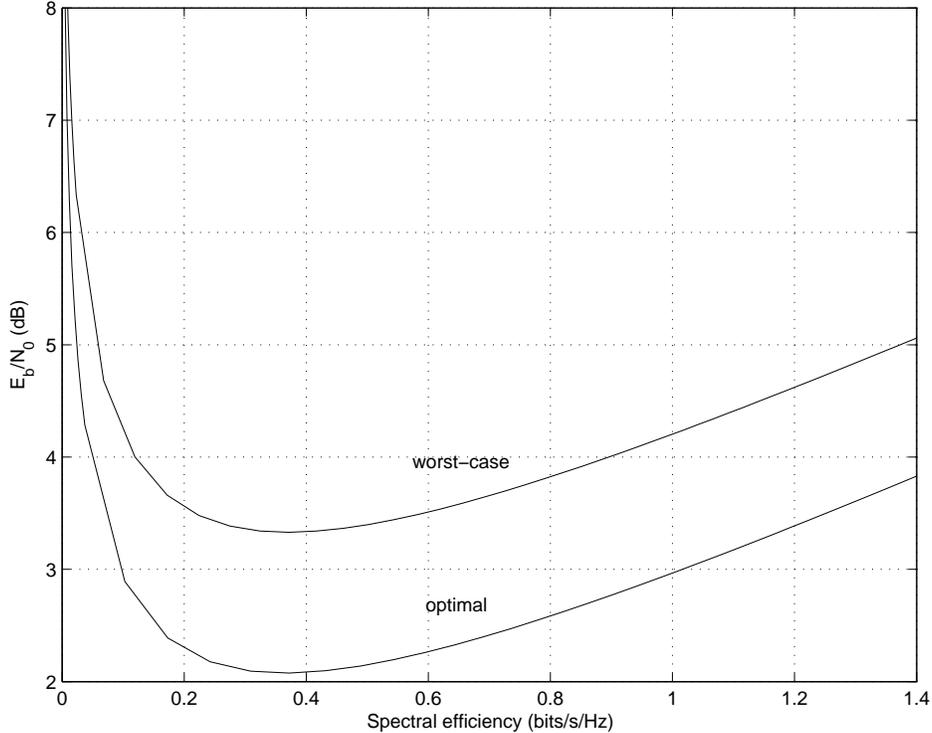}
\caption{Bit energy $\frac{E_b}{N_0}$ vs. Spectral efficiency
$\C\left(\frac{E_b}{N_0}\right)$ in the worst-case scenario and the
scenario of optimal coding-decoding under input peak-power
constraints. The block length is $m = 10$.} \label{fig:comp}
\end{center}
\end{figure}

 In the very low $\tsnr$ regime, the channel estimate
deteriorates and the performance approaches that of noncoherent
Rayleigh block fading channels. As shown in \cite{gursoy-isit}, bit
energy values required in these channels grow without bound as
$\tsnr \to 0$ and the same phenomenon is observed here as well. In
the worst-case scenario treated in Section \ref{sec:worstcase}, the
performance deterioration at very low $\tsnr$ levels is due to the
fact that poor channel estimates are assumed to be perfect. In this
section, similar observations are the result of the limitations on
the peakedness of the signal. Nevertheless, designing the
transmission and reception for channel in (\ref{eq:dataphasemodel2})
rather than that in (\ref{eq:worstcasemodel}) leads to energy gains
in the low-$\tsnr$ regime. Fig. \ref{fig:comp} provides a comparison
of the bit energy values required in the worst case scenario and the
scenario where peak power constraints are imposed and optimal
signaling and decoding is employed. In the worst-case scenario, the
channel estimate is assumed to be perfect and transmission and
reception is designed for a known channel. This is obviously a poor
assumption in the low-$\tsnr$ regime and in Fig. \ref{fig:comp} we
observe bit energy gains of approximately 1.5 dB when optimal
techniques are employed in the case of $m = 10$. Note that these
gains are achieved when the input is subject to more stringent peak
power constraints. From Fig. \ref{fig:comp}, we also conclude that
in the low-$\tsnr$ regime, the achievable rate expression in
(\ref{eq:worstcap}) is a lower bound to the peak-power limited
capacity of the channel in (\ref{eq:dataphasemodel2}). Note that
(\ref{eq:worstcap}) will eventually exceed this capacity value at
high $\tsnr$ levels as it is obtained under less strict average
power constraints.

\begin{figure}
\begin{center}
\includegraphics[width = \figsize\textwidth]{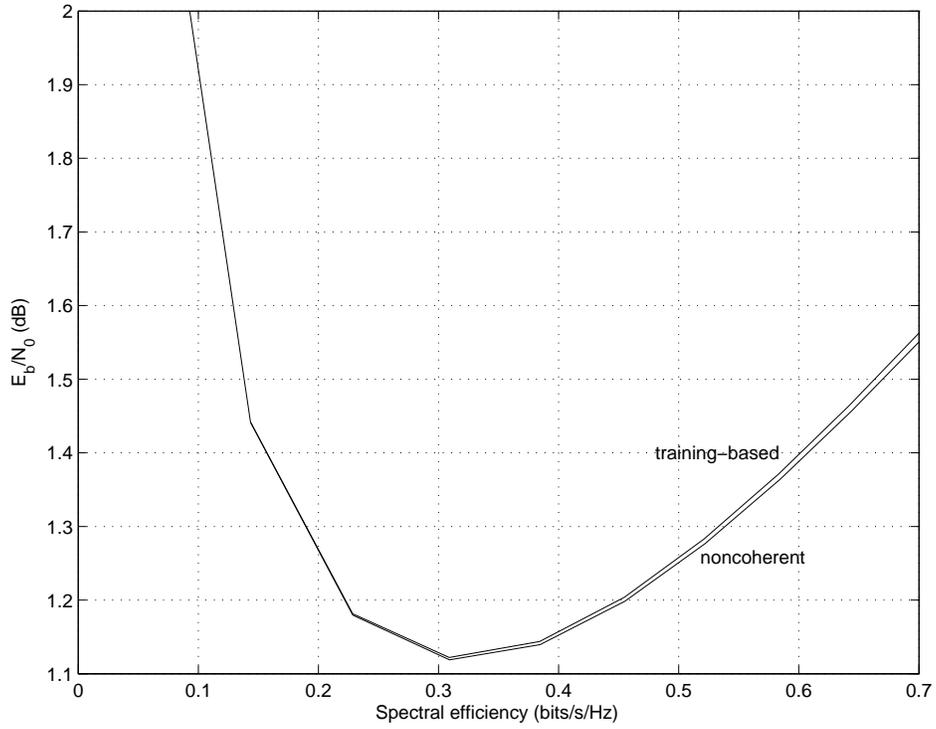}
\caption{Bit energy $\frac{E_b}{N_0}$ vs. Spectral efficiency
$\C\left(\frac{E_b}{N_0}\right)$ for training-based and noncoherent
communication systems when $m = 20$.} \label{fig:pilotvsnoncoh}
\end{center}
\end{figure}

\begin{figure}
\begin{center}
\includegraphics[width = \figsize\textwidth]{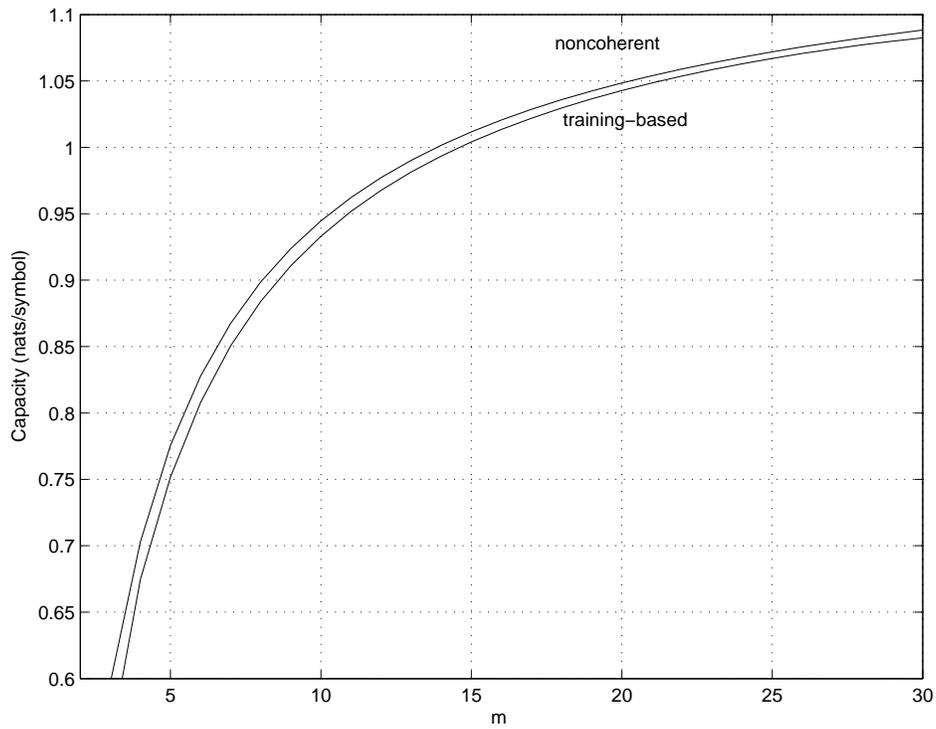}
\caption{Capacity (nats/symbol) vs. block length $m$ for
training-based and noncoherent communication systems. $\tsnr = 5$
dB.} \label{fig:m}
\end{center}
\end{figure}

In training-based systems, certain fraction of time and power which
otherwise will be used for data transmission is allocated to the
pilot symbols to facilitate channel estimation. Hence, there is a
potential for performance loss in terms of data rates. However, at
the same time, the availability of channel estimates at the receiver
tends to improve the performance. On the other hand, in noncoherent
communications, there is no attempt for channel estimation and
communication is performed over unknown channels.
The analysis presented in this paper can be applied to noncoherent
communications in a straightforward manner by choosing $\delta = 0$
and replacing $m$ in the equations by $m+1$ as no time is allocated
to pilot symbols. Hence, for instance, the discrete nature of the
optimal input under peak power constraints can easily be shown for
the noncoherent Rayleigh channel as well. However, the details of
this analysis is omitted because the discreteness results are proven
for noncoherent Rician fading channels in \cite{gursoy-isit} and for
more general noncoherent MIMO channels in \cite{Chan}. Here, we
present numerical results. Figures \ref{fig:pilotvsnoncoh} and
\ref{fig:m} compare the performances of training-based and
noncoherent communication systems. In Fig. \ref{fig:pilotvsnoncoh},
the bit energy values are plotted for both schemes when the block
length is $m = 20$. It is observed that for this relatively small
value of the block length, both schemes achieve almost the same
minimum bit energy value, and therefore, the training-based
performance is surprisingly rather close to that of the noncoherent
scheme even in the low-$\tsnr$ regime. Fig. \ref{fig:m} plots the
capacity values as a function of the block length at $\tsnr = 5$ dB.
Here, we also observe that the performance of training-based schemes
comes very close to that of the noncoherent scheme. Therefore, if
having the channel estimate reduces the complexity of the receiver
and/or pilot signals are additionally used for timing and
frequency-offset synchronization or channel equalization,
training-based schemes can be preferred over noncoherent
communications with small loss in data rates.


\subsection{Capacity with Ideal Interleaving and Per-symbol Peak Power Constraints}

Since most of the well-known codes are designed to correct errors
that occur independently from the location of other errors
\cite{Proakis}, practical communication systems employ interleavers
at the transmitters to gain protection against error bursts.
Deinterleavers are used at the receiver to reverse the interleaving
operation. In this section, we consider such systems and assume that
ideal interleaving is used so that each data symbol experiences
independent channel conditions. Pilot symbols are inserted
periodically after the interleaver. We note that a pilot-assisted
transmission with ideal interleaving is also studied in
\cite{Baltersee1} and \cite{Baltersee2} where achievable rates are
considered. Since interleaving breaks the channel correlation seen
by the data symbols, channel memory can no longer be taken advantage
of in the transmission. Hence, interleaving in general decreases the
capacity. Therefore, the capacity results in this section can also
be regarded as lower bounds on the capacity of a non-interleaved
system. On the other hand, one advantage of interleaving is the
simplification of signaling schemes.

\begin{figure}
\begin{center}
\includegraphics[width = \figsize\textwidth]{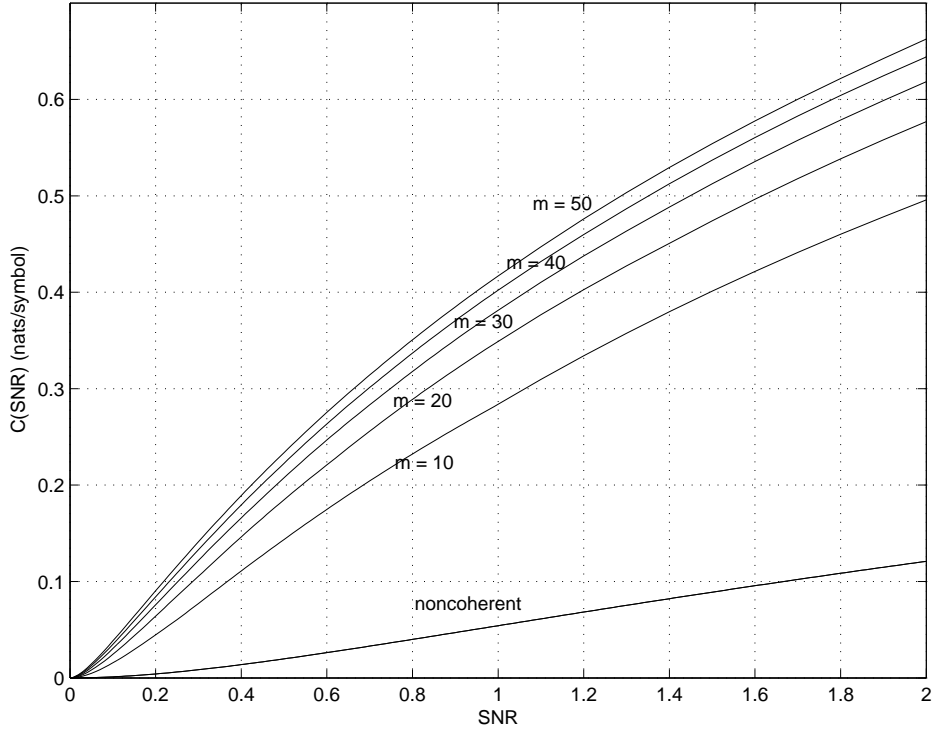}
\caption{Capacity (nats/symbol) vs. $\tsnr$ for interleaved,
training-based transmissions when block lengths are $m = 10,20,30,
40$ and $50$, and for interleaved noncoherent transmission over the
unknown Rayleigh fading channel.} \label{fig:capinterl}
\end{center}
\end{figure}

\begin{figure}
\begin{center}
\includegraphics[width = \figsize\textwidth]{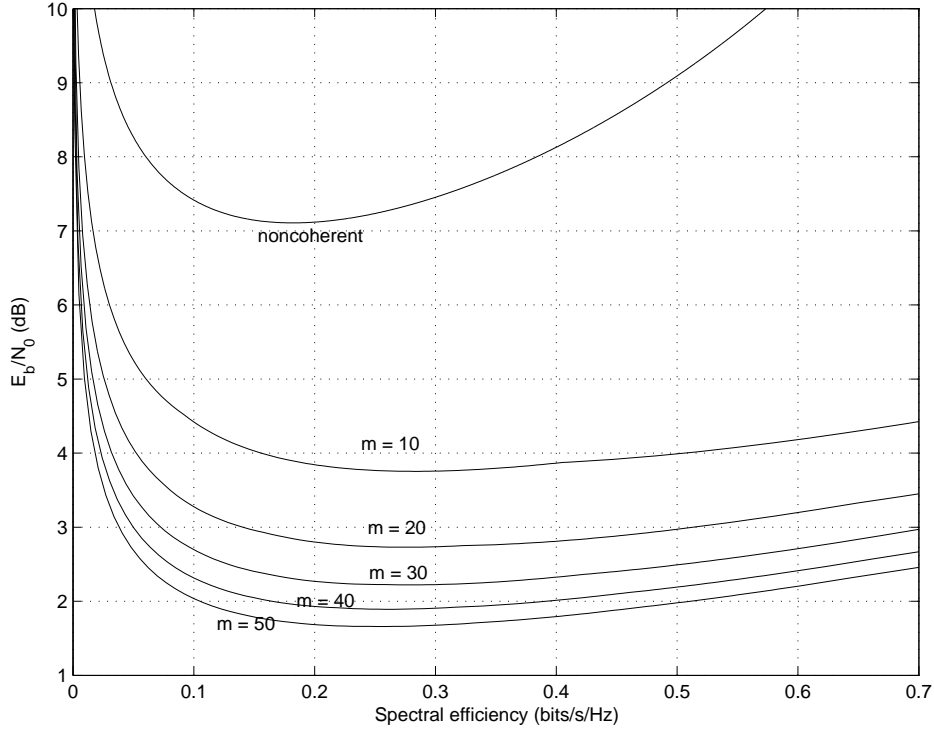}
\caption{Bit energy $\frac{E_b}{N_0}$ vs. Spectral efficiency
$\C\left(\frac{E_b}{N_0}\right)$ for interleaved, training-based
transmissions when block lengths are $m = 10,20,30, 40$ and $50$,
and for interleaved noncoherent transmission over the unknown
Rayleigh fading channel.} \label{fig:ebnointerl}
\end{center}
\end{figure}

\begin{figure}
\begin{center}
\includegraphics[width = \figsize\textwidth]{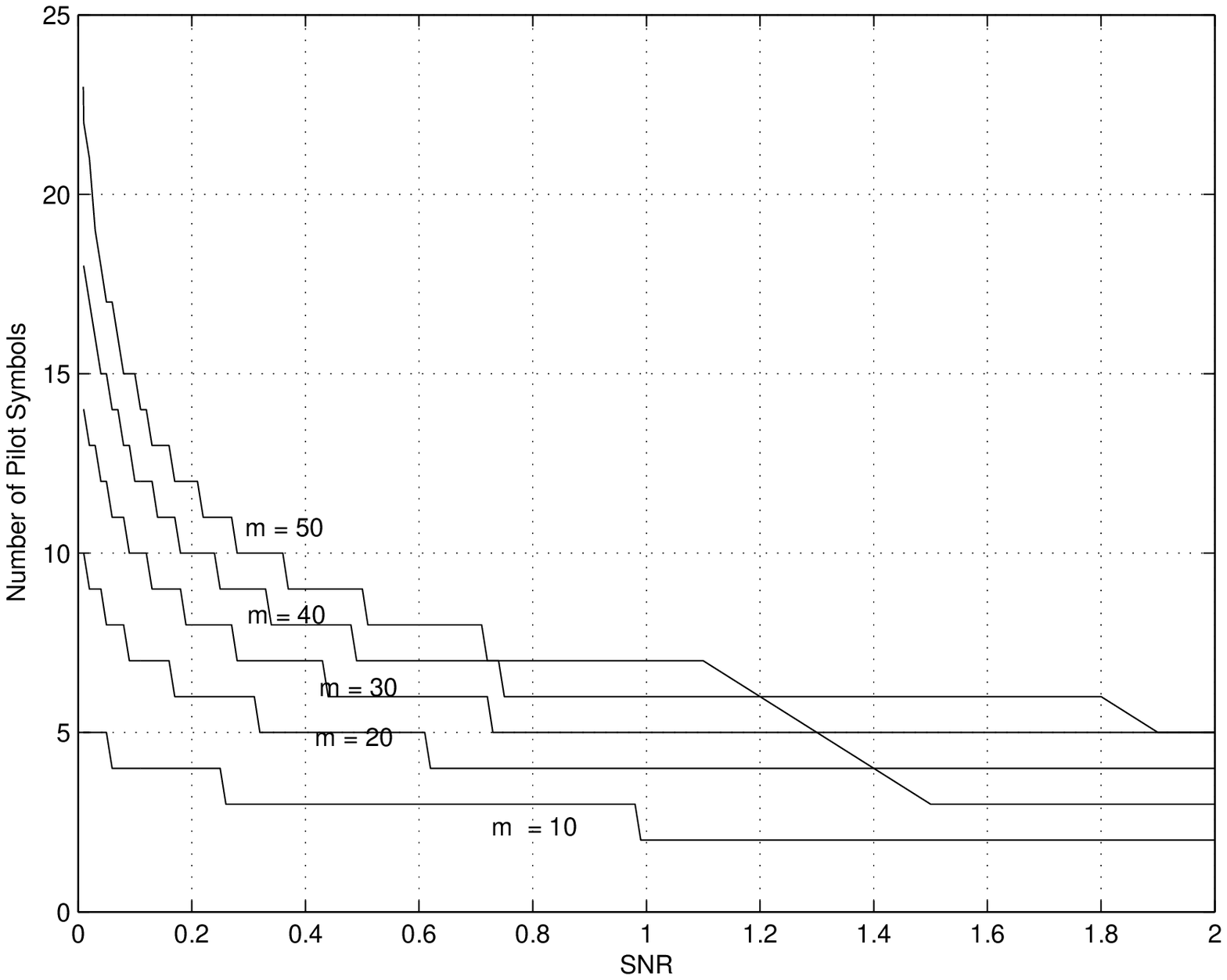}
\caption{Number of pilot symbols per block  vs. $\tsnr$ for
interleaved, training-based transmissions when block lengths are $m
= 10,20,30, 40$ and $50$.} \label{fig:noofpilotsinterl}
\end{center}
\end{figure}

We continue considering the block fading channel model. Hence, the
channel stays constant for a block of $m$ symbols. However, after
deinterleaving, the channel output can be expressed as
\begin{gather} \label{eq:dataphasemodelinterl}
y_{d,i} = \he_i x_{d,i} + \hn_i x_{d,i} + n_i \quad i = 1,2,3 \ldots
\end{gather}
Note that due to interleaving, each data symbol $x_{d,i}$ is
affected by independent and identically distributed fading
coefficients $h_i = \he _i + \hn_i$. In this section, we consider
per-symbol peak power constraints, $|x_i|^2 \stackrel{a.s.}{\le} P$
$\forall i$. Therefore, the pilot symbol power is $|x_t|^2 = P$.
Note that the use of more than one pilot may be optimal. The channel
capacity in this setting is formulated as follows:
\begin{align} \label{eq:pcapinterl}
C &= \sup_{1\le l \le m} \sup_{\substack{x_d \\ |x_d|^2
\stackrel{\text{a.s.}}{\le} P} } \frac{m-l}{m}\,I(x_d; y_d | \he)
\end{align}
where $l$ denotes the number of pilot symbols per $m$ symbols, and
\begin{align}
\he \sim \CN \left( 0, \frac{\gamma^4 l P}{\gamma^2 l  P+
N_0}\right) \text{ and } \hn \sim \CN \left( 0, \frac{\gamma^2 N_0
}{\gamma^2 l P + N_0}\right). \nonumber
\end{align}
The inner maximization in (\ref{eq:pcapinterl}) becomes a special
case of the inner maximization in (\ref{eq:pcap}) when we reduce the
dimensionality of the optimization problem in
(\ref{eq:pcap})\footnote{Note that the input constraints, error
variances, and the constants multiplying the mutual information
expressions will be different in the specialized case of
(\ref{eq:pcap}) and in (\ref{eq:pcapinterl}). But, the general
structures of the two optimization problems are the same.} by
choosing $m= 2$. Therefore, the results on the structure of the
capacity-achieving input immediately apply to the setting we
consider in this section. The optimal input has a uniformly
distributed phase. With this characterization, the capacity is
\begin{gather} \label{eq:pcaprinterl}
C = \sup_{1\le l \le m}
 \sup_{\substack{F_r \\ r \stackrel{\text{a.s.}}{\le} \sqrt{L}} }
\frac{m-l}{m} \, I(F_r | \he)
\end{gather}
where
\begin{align} \label{eq:mutualinfoisotropinterl}
I(F_r | \he) = -E_{\K, r} \left\{ \int_0^\infty f_{R|r,\K}(R|r,\K)
\log \,g(R,F_r, \K ) \, dR \right\} - E_r \{\log(1+r^2)\} - 1
\end{align}
where
\begin{gather}
f_{R|r,\K}(R|r,\K) =  \frac{e^{- \frac{R + \K r^2}{1+r^2}}}{1+r^2}
I_0\left(\frac{2\sqrt{\K R}\,r}{1+r^2} \right), \label{eq:fRrinterl}
\\
g(R,F_r, \K) = \int_0^\infty f_{R|r,\K}(R|r,\K) \, dF_r,
\label{eq:gRinterl}
\end{gather}
and, $R = \frac{|y_d|^2}{N_0}$, $r = \frac{\tgamma
|x_d|}{\sqrt{N_0}}$, $\K = \frac{|\he|^2}{\tgamma^2}$, $\tgamma^2 =
\frac{\gamma^2 N_0 }{\gamma^2 l P + N_0}$, and $L =
\frac{\gamma^2P/N_0}{l \gamma^2P/N_0 + 1} = \frac{\tsnr}{l \tsnr +
1}$. Note that $\K$ is an exponential random variable with mean
$E\{\K\} = \frac{E\{|\he^2|\}}{\tgamma^2} = \frac{l \gamma^2 P}{N_0}
= l \tsnr$. Since the inner maximization in (\ref{eq:pcaprinterl})
is a special case of that in (\ref{eq:pcapr}), we immediately have
the following result.
\begin{theo:discreteinterl} \label{theo:discreteinterl}
Fix the value of $1 \le l \le m$. The input distribution that
maximizes the mutual information $I(F_r|\he)$ in
(\ref{eq:pcaprinterl}) is discrete with a finite number of mass
points.
\end{theo:discreteinterl}
Next, we present numerical results. Fig. \ref{fig:capinterl} plots,
for different values of the block lengths, the capacity curves as a
function of $\tsnr$ for training-based schemes. We observe that the
capacity values increase with the block length even though the
channel in (\ref{eq:dataphasemodelinterl}) is memoryless. This
performance gain should be attributed to the fact that the channel
estimate improves with increasing block length. Fig.
\ref{fig:capinterl} also plots the capacity of the interleaved
noncoherent communications in which no attempt is made to learn the
channel. From the comparison of the capacity curves, we observe that
training significantly enhances the data rates when data symbols are
interleaved at the transmitter. In Fig. \ref{fig:ebnointerl}, bit
energy curves as a function of the spectral efficiency are plotted.
Again, we see that training-based schemes perform much better in
terms of energy efficiency than the noncoherent scheme. In all
cases, the minimum bit energy is achieved at a nonzero spectral
efficiency level below which one should not operate. The bit energy
requirement increases without bound as spectral efficiency decreases
to zero. When we compare Figs. \ref{fig:ebno} and
\ref{fig:ebnointerl}, we note that while simplifying the system
design, interleaving also incurs a penalty in energy efficiency.
Finally, in Fig. \ref{fig:noofpilotsinterl}, we provide the optimal
resource allocations by plotting the optimal number of pilot symbols
per block as a function of $\tsnr$ for different block length
values. We realize that optimal number of pilots tends to increase
as $\tsnr$ decreases and approaches $m/2$. Hence, as in Section
\ref{subsec:worstcaseavgpower}, asymptotically half of the available
power in each block should be allocated to the training symbols.


\subsection{Achievable Rates and Bit Energies of On-Off Keying}

In this section, we relax the input constraints and assume that the
input is subject to an average power constraint
\begin{gather}
E\{\|\x\|^2\} \le mP. \label{eq:avgpowerook}
\end{gather}
We consider the channel model (\ref{eq:dataphasemodel2}) where there
is no interleaving. Akin to Section \ref{subsec:worstcaseflash}, our
goal is to obtain the attainable bit energy levels when signals with
high peak-to-average power ratios are employed. As before, single
pilot symbol with power $|x_t|^2 = \delta mP$ is used and hence the
data vector is subject to $E\{\|\x_d\|^2\} \le (1-\delta)mP$. The
data vector is again assumed to have an isotropically distributed
directional vector $\bv$, and hence $\x_d = \|\x_d\| \bv$. We
further assume that the on-off keying is used for magnitude
modulation and therefore
\begin{align}
\frac{1}{\sqrt{m}}\|\x_d\| = \left \{
\begin{array}{ll}
A & \text{with prob.} \quad p_0
\\
0 & \text{with prob.} \quad 1-p_0
\end{array}\right.
\end{align}
where $A$ is a fixed magnitude level that does not vary with the
power $P$. In order to satisfy the average power constraint we
should have
\begin{gather}
A^2 p_0 = (1 - \delta)P \Rightarrow p_0 = \frac{(1-\delta)P}{A^2}
\end{gather}
Therefore, in this signaling scheme, the peak power of the
transmitted data signal is kept constant while its probability
vanishes as $P \to 0$. Hence, while the peak power is fixed, the
peak-to-average power ratio grows without bound as $P \to 0$.
Similarly as before, we define $r = \frac{\tgamma
\|\x_d\|}{\sqrt{N_0}}$. With this definition, the distribution of
$r$ is
\begin{align}
r = \left \{
\begin{array}{ll}
\frac{\gamma \sqrt{m} A}{\sqrt{\gamma^2 \delta m P + N_0}} &
\text{with prob.} \quad p_0 = \frac{(1-\delta)P}{A^2}
\\
0 & \text{with prob.} \quad 1-p_0
\end{array}\right..
\end{align}
We further define $\nu = \frac{A^2 \gamma^2 }{(1-\delta) N_0}$ which
does not depend on $P$, and $\tsnr = \frac{\gamma^2 P}{N_0}$. Now,
we can write
\begin{align} \label{eq:ookdistr}
r = \left \{
\begin{array}{ll}
r_0 = \frac{\sqrt{(1-\delta)m \nu}}{\sqrt{\delta m \tsnr + 1}} &
\text{with prob.} \quad p_0 = \frac{\tsnr}{\nu}
\\
0 & \text{with prob.} \quad 1-p_0
\end{array}\right..
\end{align}
For a given value of $\delta$, the mutual information achieved by
the isotropically distributed directional vector $\bv$ and $r$ whose
distribution is given in (\ref{eq:ookdistr}) is
\begin{align} \label{eq:mutualinfoisotropook}
I_{ook} = -E_{\K} \left\{ \int_0^\infty f_{R|\K}(R|\K) \log
\left(\frac{(m-2)!}{R^{m-2}} f_{R|\K}(R|\K)\right) \, dR \right\} -
p_0 \log(1+r_0^2) - (m-1)
\end{align}
where $f_{R|\K}(R|\K) = (1-p_0)f_{R|r,\K}(R|r=0,\K) + p_0
f_{R|r,\K}(R|r=r_0,\K)$ and $f_{R|r,\K}(R|r,\K)$ is given in
(\ref{eq:fRr}). Note that $K$ is an exponential random variable with
mean $E\{\K\} = \frac{\delta m \gamma^2 P}{N_0} = \delta m \tsnr$.
Next, we obtain the bit energy required for reliable communications
with OOK as $\tsnr \to 0$.
\begin{theo:bitenergyook} \label{theo:bitenergyook}
Assume that the normalized input magnitude distribution is given by
(\ref{eq:ookdistr}). For a given value of $\delta \in (0,1)$, the
normalized bit energy required by this input as $P \to 0$ is
\begin{gather}
\left.\frac{E_{b,ook}}{N_0}\right|_{I_{ook} = 0} = \lim_{\tsnr \to
0} \frac{m\tsnr}{I_{ook}(\tsnr)}\log2 =
\frac{m\log2}{\dot{I}_{ook}(0)} = \frac{\log2}{(1-\delta) -
\frac{1}{m\nu}\log(1+(1-\delta)m\nu)}.
\end{gather}
\end{theo:bitenergyook}
\vspace{0.2cm}\emph{Proof}: As in the proof of Theorem
\ref{theo:bitenergysinglemass}, we apply a change of variables and
express the mutual information as
\begin{gather} \label{eq:mutualinfoisotropookcov}
I_{ook} = -E_{\K} \left\{ \int_0^\infty f_{R|\K}(R|\K\delta m\tsnr)
\log \left(\frac{(m-2)!}{R^{m-2}} f_{R|\K}(R|\K\delta m
\tsnr)\right) \, dR \right\} - p_0 \log(1+r_0^2) - (m\!-\!1)
\end{gather}
where $\K$ is now an exponential random variable with mean $E\{\K\}
= 1$. It can be easily seen that
\begin{gather} \label{eq:pologderiv}
\left.\frac{\partial}{\partial \tsnr}p_0 \log(1+r_0^2)\right|_{\tsnr
= 0} = \frac{1}{\nu} \log(1 + (1-\delta)m\nu).
\end{gather}
We can also show that
\begin{gather}
\left.\frac{\partial}{\partial \tsnr} f_{R|\K}(R|\K \delta m
\tsnr)\right|_{\tsnr = 0} = -\frac{1}{\nu} f_{R|r,\K}(R| r = 0, K =
0) + \frac{1}{\nu} f_{R|r,\K}(R| r = \sqrt{(1-\delta)m\nu}, K = 0).
\label{eq:frKderiv}
\end{gather}
Using (\ref{eq:frKderiv}), we can prove that the derivative of the
first term on the right-hand side of
(\ref{eq:mutualinfoisotropookcov}) with respect to $\tsnr$ at $\tsnr
= 0$ is $(1-\delta)m$. Combining this result with
(\ref{eq:pologderiv}), we arrive to
\begin{gather}
\dot{I}_{ook}(0) = (1-\delta)m - \frac{1}{\nu} \log(1 +
(1-\delta)m\nu)
\end{gather}
which concludes the proof. \hfill $\square$

Theorem \ref{theo:bitenergyook} shows that unlike previously treated
cases, reliable communications with OOK modulation with fixed peak
power requires finite bit energy as $P \to 0$. Hence, OOK provides
significant improvements in energy efficiency in the low-$\tsnr$
regime at the cost of high peak-to-average power ratio. Since $\nu =
\frac{A^2 \gamma^2 }{(1-\delta) N_0}$, we can also express the
asymptotic bit energy level as
\begin{gather} \label{eq:bitenergyook2}
\left.\frac{E_{b,ook}}{N_0}\right|_{I_{ook} = 0} =
\frac{\log2}{(1-\delta)\left(1 - \frac{1}{m \frac{A^2
\gamma^2}{N_0}}\log\left(1+m\frac{A^2 \gamma^2}{N_0}\right)\right)}.
\end{gather}
It has been shown in \cite{gursoy-isit} that, if noncoherent
communications with no channel estimation is performed and the input
is subject to $E\{\|\x|^2\} \le mP$ and $\|\x\|^2
\stackrel{\text{a.s.}}{\le} mA$, then optimal signaling requires the
following bit energy value as $P \to 0$:
\begin{gather} \label{eq:bitenergyfixedpeak}
\left.\frac{E_{b, noncoh}}{N_0}\right|_{\C = 0} =
\frac{\log2}{\left(1 - \frac{1}{m \frac{A^2
\gamma^2}{N_0}}\log\left(1+m\frac{A^2 \gamma^2}{N_0}\right)\right)}.
\end{gather}
We note that similar results for fading channels with memory are
obtained in \cite{Sethuraman} through the analysis of capacity per
unit cost. Comparing (\ref{eq:bitenergyook2}) and
(\ref{eq:bitenergyfixedpeak}), we find that training-based schemes
suffer an energy penalty due to the presence of the term
$1/(1-\delta)$ and this penalty vanishes if $\delta \to 0$.
Therefore, if OOK with fixed power is employed, the power of the
training symbols should be decreased to zero as $P \to 0$ to match
the  noncoherent performance. This power allocation policy is in
stark contrast to the results in the previous sections. Note that as
$\tsnr$ decreases, data transmission occurs extremely infrequently.
In such a case, performing channel estimation all the time for each
$m$-block irrespective of whether or not data transmission takes
place is not an good design choice. Hence, a gradual decrease in the
power allocated to training should also be intuitively expected. We
further remark that as $m \to \infty$ and $\delta \to 0$,
$\left.\frac{E_{b,ook}}{N_0}\right|_{I_{ook} = 0} \to -1.59$ dB.

\begin{figure}
\begin{center}
\includegraphics[width = \figsize\textwidth]{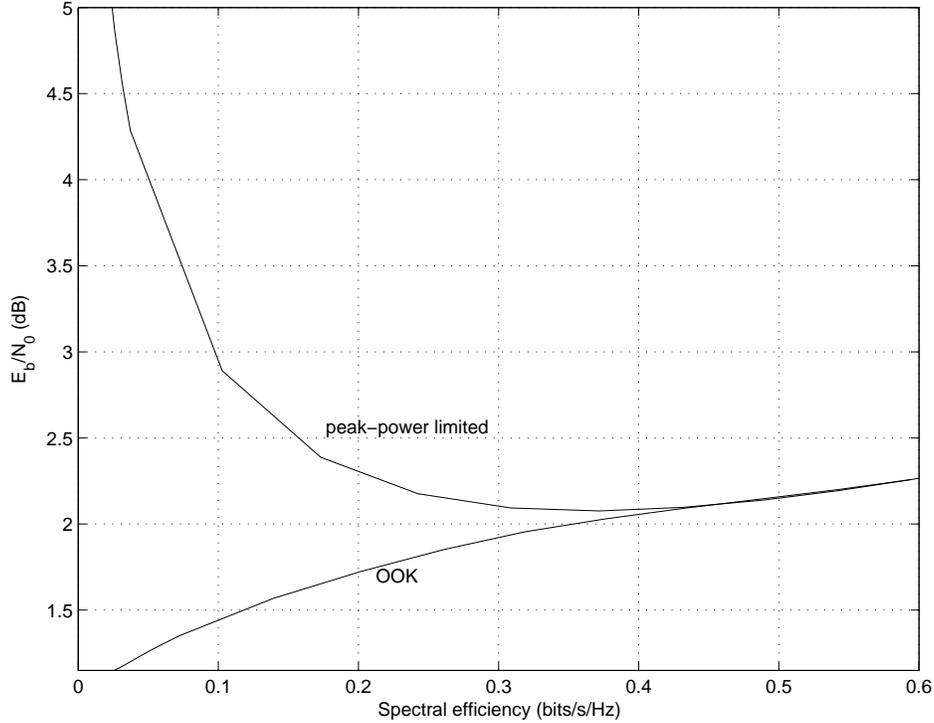}
\caption{Bit energy $\frac{E_b}{N_0}$ vs. Spectral efficiency
$\C\left(\frac{E_b}{N_0}\right)$ for training-based OOK signaling
and training-based optimal signaling under input peak power
constraints. The block length is $m = 10$.} \label{fig:flashinterl}
\end{center}
\end{figure}

Fig. \ref{fig:flashinterl} plots the bit energy levels as a function
of spectral efficiency for training-based OOK with fixed peak power
and for training-based optimal signaling under input peak power
constraints in the form $\|\x\|^2
\stackrel{\text{a.s.}}{\le}(1-\delta)mP$. In this figure, the block
length is $m = 10$, and for OOK, $\nu = 1$. As predicted, below the
spectral efficiency of approximately 0.4 bits/s/Hz, OOK provides
better energy efficiency. The bit energy requirements of OOK
decreases as spectral efficiency decreases as opposed to the
behavior presented in the peak-power-limited case. Numerical
analysis have also shown that the fraction of power allocated to
training, $\delta$, in OOK decreases as $\tsnr$ decreases,
conforming with the discussion in the previous paragraph.


\section{Conclusion} \label{sec:conclusion}

In this paper, we have studied the energy efficiency and capacity of
training-based communication schemes employed for the transmission
of information over a-priori unknown Rayleigh block fading channels.
We have initially considered the worst-case scenario in which the
product of the estimate error and transmitted signal is assumed to
be Gaussian noise. The capacity expression obtained under this
assumption is a lower bound to the true capacity of the channel, and
provides the achievable rates when the communication system is
designed as if the channel estimate were perfect. We have
investigated the bit energy levels required for reliable
communications and quantified the penalty in energy efficiency
incurred due to regarding the imperfect channel estimate as perfect
in the low-$\tsnr$ regime. We have shown that the bit energy
requirements grow without bound as $\tsnr \to 0$ regardless of the
size of the block length $m$. Hence, the minimum bit energy is
achieved at a nonzero $\tsnr$ value below which one should not
operate under the aforementioned assumptions. We have also shown
that approaching the minimum bit energy level of $-1.59$ dB is
extremely slow in terms of block length as $m \to \infty$. Similar
results are obtained if peak power limitations are imposed on
training symbols. We have also investigated flash training and
transmission schemes to improve the energy efficiency at low $\tsnr$
levels. We have shown that in order for the bit energy requirement
not to grow as $\tsnr \to 0$, the duty cycle in flash transmission
should vanish linearly with decreasing $\tsnr$.

Next, we have analyzed the capacity and energy efficiency of
training-based schemes when the input is subject to peak power
constraints. We have characterized that the capacity-achieving input
has a discrete magnitude and an isotropically distributed unit
directional vector. Using this characterization, we have obtained
the capacity expressions, optimal training power allocations, and
bit energy levels required for reliable communications. We have
noted that at low $\tsnr$s, the optimal input magnitude is fixed at
a constant level. Due to the presence of the peak power constraints,
the bit energy requirements are again shown to increase without
bound as $\tsnr \to 0$. However, we have seen that gains in energy
efficiency are obtained when optimal signaling and decoding are
employed. We have compared the performances of training-based and
noncoherent transmission schemes. Although training-based schemes
dedicate certain amount of time and power to training symbols and as
a result are expected to suffer in terms of data rates, we have
observed that the performance loss is small even at relatively small
block lengths and small $\tsnr$ levels. We have also considered the
case in which interleaving used at the transmitter for protection
against error bursts and per-symbol peak power constraints are
imposed. We have obtained the channel capacity, optimal training
duration, and analyzed the energy efficiency. In this case, training
is shown to improve the performance with respect to noncoherent
communications. We have also investigated the improvements in energy
efficiency in the low-$\tsnr$ regime if OOK with fixed peak power
and vanishing duty cycle is employed at the transmitter. Finally, we
note that this work has primarily focused on block fading channels.
Recently, we in \cite{ciss06}, \cite{sami1} and \cite{sami2} have
considered more general fading processes with memory. Since the
exact capacity is rather difficult to obtain in such cases,
achievable rate expressions are analyzed, and subsequently energy
efficiency and optimal resource allocations are studied.

\newpage
\appendix

\subsection{Derivation of the Mutual Information Expression in Theorem \ref{prop:eqvmutual}} \label{app:eqvmutual}

The input-output mutual information expression for channel
(\ref{eq:dataphasemodel2}) is

\begin{align}
I(\x_d ; \y_d | \he) &= E_{\he} E_{\x_d} \int f_{\y|\x_d, \he}(\y |
\x_d, \he) \log \frac{f_{\y|\x_d, \he}(\y | \x_d, \he)}{f_{\y|
\he}(\y |\he)} \, d \y \label{eq:mutualinfoapp1}
\\
&=-E_{\he} E_{\x_d} \int f_{\y|\x_d, \he}(\y | \x_d, \he) \log
f_{\y| \he}(\y |\he) \, d \y - E_{\x_d}\{\log (\pi^{m-1} N_0^{m-2}
e^{m-1} (\tgamma^2\|\x_d\|^2 + N_0))\}\label{eq:mutualinfoapp2}
\\
&=-E_{\he} E_{\x_d} \int f_{\y|\x_d, \he}(\y | \x_d, \he) \log
f_{\y| \he}(\y |\he) \, d \y - E_{r}\{\log (1+r^2)\} - \log
(\pi^{m-1} N_0^{m-1} ) -(m-1) \label{eq:mutualinfoapp3}
\end{align}
Note that the second part of (\ref{eq:mutualinfoapp2}) is the
conditional differential entropy of $\y$ given $\x_d$ and $\he$.
(\ref{eq:mutualinfoapp3}) follows from the definition $r =
\frac{\tgamma \|x_d\|}{\sqrt{N_0}}$. The main difficulty is to
simplify
\begin{gather}
\chi(\x_d, \he) = \int f_{\y|\x_d, \he}(\y | \x_d, \he) \log f_{\y|
\he}(\y |\he) \, d \y
\end{gather}
which, in general, is an $2(m-1)$-fold integral. Note that
\begin{gather}
f_{\y| \he}(\y |\he) = \int f_{\y|\x_d, \he}(\y | \x_d, \he) \,
dF_{\x_d}.
\end{gather}
Using the facts that $f_{\y|\x}(\Phi \y|\Phi \x_d, \he) =
f_{\y|\x_d, \he}(\y | \x_d, \he)$ and input has circular symmetry,
we can easily see that for any fixed unitary matrix $\Phi$
\begin{gather}
f_{\y| \he}(\Phi\y |\he) = f_{\y| \he}(\y |\he) = f_{\y| \he}(\|\y\|
|\he)
\\ \intertext{and hence}
\chi(\Phi \x_d, \he) = \chi(\x_d, \he) = \chi(\|\x_d\|, \he).
\end{gather}
Therefore, $f_{\y| \he}(\y |\he)$ and $\chi(\x_d, \he)$ are
circularly-symmetric functions depending only on $\|\y\|$ and
$\|\x_d\|$, respectively. Noting that
\begin{gather}
(\tgamma^2 \x_d \x_d^\dagger + N_0 I)^{-1} = \frac{\I}{N_0} -
\frac{\tgamma^2 \x_d \x_d^\dd}{N_0 (\tgamma^2 \|\x_d\|^2 + N_0)},
\end{gather}
and defining $\x_d = \|\x_d\|\bv$ and $\y = \|\y\| \bw$, we can,
after some algebraic steps, rewrite the conditional density function
in (\ref{eq:condpdf}) as
\begin{gather} \label{eq:condpdf2}
f_{\y|\x_d, \he}(\y | \x_d, \he) =
\frac{\exp\left(-\frac{\|\y\|^2}{N_0}- \frac{|\he|^2
\|\x_d\|^2}{\tgamma^2 \|\x_d\|^2 + N_0} + \frac{\tgamma^2 \|\x_d\|^2
\|\y\|^2 |\bw^\dd \bv|^2}{N_0(\tgamma^2 \|\x_d\|^2 + N_0)} + \frac{2
\|\x_d\| \|\y\||\he| \, \Re(e^{j\theta_{\he}} \bw^\dd
\bv)}{\tgamma^2 \|\x_d\|^2 + N_0}\right)} {\pi^{m-1}
N_0^{m-2}(\tgamma^2 \|\x_d\|^2 + N_0)}
\end{gather}
where $\Re(z)$ denotes the real part of the complex number $z$, and
$\theta_{\he}$ is the phase of $\he$. The usefulness of
(\ref{eq:condpdf2}) comes from the property that the magnitude
$\|\x_d\|$ and the directional unit vector $\bv$ are separated. We
know from Theorem \ref{prop:blisotropic} that $\bv$ is isotropically
distributed and independent of $\|\x_d\|$. Hence, we now have
\begin{gather}
f_{\y| \he}(\y |\he) = \int \frac{\exp\left(-\frac{\|\y\|^2}{N_0}-
\frac{|\he|^2 \|\x_d\|^2}{\tgamma^2 \|\x_d\|^2 + N_0} +
\frac{\tgamma^2 \|\x_d\|^2 \|\y\|^2 |\bw^\dd \bv|^2}{N_0(\tgamma^2
\|\x_d\|^2 + N_0)} + \frac{2 \|\x_d\| \|\y\||\he| \,
\Re(e^{j\theta_{\he}} \bw^\dd \bv)}{\tgamma^2 \|\x_d\|^2 +
N_0}\right)} {\pi^{m-1} N_0^{m-2}(\tgamma^2 \|\x_d\|^2 + N_0)} \,
f_\bv(\bv) dF_{\|\x_d\|}.
\end{gather}
where $f_\bv$ is the probability density function of $\bv$. Since
$f_{\y| \he}$ is a function of only $\|\y\|$, we can, without loss
of generality, assume that $\bw^\dd = [1, 0, 0, \ldots, 0]$. In such
a case,
\begin{gather}
f_{\y| \he}(\y |\he) = \int \frac{\exp\left(-\frac{\|\y\|^2}{N_0}-
\frac{|\he|^2 \|\x_d\|^2}{\tgamma^2 \|\x_d\|^2 + N_0} +
\frac{\tgamma^2 \|\x_d\|^2 \|\y\|^2 |v_1|^2}{N_0(\tgamma^2
\|\x_d\|^2 + N_0)} + \frac{2 \|\x_d\| \|\y\||\he| \,
\Re(e^{j\theta_{\he}} v_1)}{\tgamma^2 \|\x_d\|^2 + N_0}\right)}
{\pi^{m-1} N_0^{m-2}(\tgamma^2 \|\x_d\|^2 + N_0)} \, f_{v_1}(v_1)
dF_{\|\x_d\|}.
\end{gather}
where $v_1$ is the first component of $\bv$ and $f_{v_1}$ is the
corresponding density function. From \cite{Marzetta}, we have for $m
\ge 3$
\begin{gather}
f_{v_1}(v_1) = \frac{1}{2\pi} 2(m-2)(1-|v_1|^2)^{m-3} \quad |v_1|
\le 1.
\end{gather}
Hence, $v_1$ has a uniform phase and a magnitude whose density
function is
\begin{gather}
f_{|v_1|}(|v_1|) = 2(m-2)|v_1| (1-|v_1|^2)^{m-3}.
\end{gather}
Note that if $m =2$, then $\x_d$ is one-dimensional and hence $\x_d
= \|\x_d\| \bv = \|\x_d\| e^{\theta_{\x_d}}$. Therefore, in this
case, $|\bv| = |v_1| = 1$ with probability one. Using these facts
and defining $r = \frac{\tgamma \|\x_d\|}{\sqrt{N_0}}$, $R =
\frac{\|\y\|^2}{N_0}$, $\K = \frac{|\he|^2}{\tgamma^2}$, and $a =
|v_1|^2$, we obtain
\begin{align}
 f_{\y| \he}(\y |\he) &= \left\{
\begin{array}{ll}
 \int_0^\infty d F_r \,\frac{(m-2)e^{-R - \frac{\K
r^2}{1+r^2}}}{\pi^{m-1}N_0^{m-1}(1+r^2)} \int_0^1  \,(1-a)^{m-3}
e^{\frac{ar^2R}{1+r^2}} \, I_0\left(\frac{2\sqrt{\K R} \, r
\sqrt{a}}{1+r^2} \right) \,da & m\ge 3
\\
 \int_0^\infty d F_r \,\frac{e^{- \frac{R + \K
r^2}{1+r^2}}}{\pi^{m-1}N_0^{m-1}(1+r^2)} I_0\left(\frac{2\sqrt{\K R}
\, r }{1+r^2} \right) & m = 2
\end{array}\right.
\\
&= \frac{g(R,F_r,\K)}{\pi^{m-1}N_0^{m-1}}
\end{align}
where $g(R,F_r,\K)$ is defined in (\ref{eq:gR}). Therefore, we have
\begin{align}
\chi(\x_d, \he) &= \int f_{\y|\x_d, \he}(\y | \x_d, \he) \log f_{\y|
\he}(\y |\he) \, d \y
\\
&= E_{\y|\x_d, \he}\{\log f_{\y| \he}(\y |\he) \, d \y\}
\\
&= E_{R|r, \K}\{\log
\left(\frac{g(R,F_r,\K)}{\pi^{m-1}N_0^{m-1}}\right) \}
\\
&= -\log (\pi^{m-1}N_0^{m-1} ) + E_{R|r, \K}\{\log g(R,F_r,\K) \}
\\
&= -\log (\pi^{m-1}N_0^{m-1} ) + \int_0^\infty f_{R|r,\K}(R|r,\K)
\log g(R,F_r,\K) \, dR \label{eq:chi}
\end{align}
where $f_{R|r,\K}(R|r,\K)$ is the conditional density function of
$R$ given $r$ and $\K$. Combining (\ref{eq:mutualinfoapp3}) and
(\ref{eq:chi}), we get
\begin{align}
I(\x_d; \y_d | \he ) = -E_{\K, r} \left\{ \int_0^\infty
f_{R|r,\K}(R|r,\K) \log \,g(R,F_r, \K ) \, dR \right\} - E_r
\{\log(1+r^2)\} - (m-1)
\end{align}
which is the mutual information expression provided in Theorem
\ref{prop:eqvmutual}. Proof will be completed by showing that
$f_{R|r,\K}(R|r,\K)$ has the expression given in (\ref{eq:fRr}).
From the previous development, we can easily verify that
\begin{gather}
f_{R_1,\ldots,R_{m-1}|r,\K}(R_1,\ldots,R_{m-1}|r,\K) = \left\{
\begin{array}{ll}
\frac{(m-2)e^{-R - \frac{\K r^2}{1+r^2}}}{(1+r^2)} \int_0^1
\,(1-a)^{m-3} e^{\frac{ar^2R}{1+r^2}} \, I_0\left(\frac{2\sqrt{\K
R}\, r \sqrt{a}}{1+r^2} \right) \,da & m \ge 3
\\
\frac{e^{- \frac{R  + \K r^2}{1+r^2}}}{(1+r^2)}
I_0\left(\frac{2\sqrt{\K R}\, r }{1+r^2} \right) \,da & m = 2
\end{array}\right.
\end{gather}
where $f_{R_1,\ldots,R_{m-1}|r,\K}$ is the conditional joint density
function of $R_1,\ldots,R_{m-1}$ given $r, \K$. Note that we above
have defined $R_i = \frac{|y_i|^2}{N_0}$ and hence $R =
\frac{\|\y\|^2}{N_0} = R_1 + R_2 + \ldots + R_{m-1}$. Note that the
joint probability density function depends on the sum $R$. We have
the following relationship
\begin{align}
\int_0^\infty f_R(R|r,\K) \, dR &= \int
f_{R_1,\ldots,R_{m-1}|r,\K}(R_1,\ldots,R_{m-1}|r,\K) \, dR_1 \ldots
dR_{m-1} \label{eq:int1}
\\ &= \int dR_2 \ldots
dR_{m-1} \, \int_{R_2 + \ldots + R_{m-1}}^\infty
f_{R_1,\ldots,R_{m-1}|r,\K}(R|r,\K) \,dR \label{eq:int2}
\\
&= \int dR_3 \ldots dR_{m-1} \, \int_{R_3 + \ldots + R_{m-1}}^\infty
f_{R_1,\ldots,R_{m-1}|r,\K}(R|r,\K) \,dR
\int_0^{R-(R_3+\ldots+R_{m-1})} dR_2 \label{eq:int3}
\\
&= \int dR_3 \ldots dR_{m-1} \, \int_{R_3 + \ldots + R_{m-1}}^\infty
(R-(R_3+\ldots+R_{m-1}))f_{R_1,\ldots,R_{m-1}|r,\K}(R|r,\K) \,dR
\label{eq:int4}
\\
&= \int_0^\infty \frac{R^{m-2}}{(m-2)!}
f_{R_1,\ldots,R_{m-1}|r,\K}(R|r,\K) \, dR. \label{eq:int5}
\end{align}
(\ref{eq:int2}) follows by applying the change of variables with $R
= R_1 + R_2 + \ldots + R_{m-1}$ in the integral with respect to
$R_1$. (\ref{eq:int3}) is obtained by interchanging the integrals
with respect to $R_2$ and $R$. (\ref{eq:int4}) follows by evaluating
the rightmost integral in (\ref{eq:int3}). Finally, (\ref{eq:int5})
is obtained through the repeated application of this procedure. From
(\ref{eq:int5}), we have
\begin{align} \label{eq:fRrapp}
f_R(R|r,\K) &= \frac{R^{m-2}}{(m-2)!}
f_{R_1,\ldots,R_{m-1}|r,\K}(R|r,\K)
\\
&= \left\{
\begin{array}{ll}
\frac{R^{m-2}}{(m-3)!} \frac{e^{-R - \frac{\K r^2}{1+r^2}}}{(1+r^2)}
\int_0^1  \,(1-a)^{m-3} e^{\frac{ar^2R}{1+r}} \,
I_0\left(\frac{2\sqrt{\K R} \, r \sqrt{a}}{1+r^2} \right) \,da & m
\ge 3
\\
\frac{e^{- \frac{R + \K r^2}{1+r^2}}}{(1+r^2)}
I_0\left(\frac{2\sqrt{\K R} \, r }{1+r^2} \right) \,da & m = 2
\end{array} \right.
\end{align}
which is the same as the expression in (\ref{eq:fRr}).

\subsection{Proof of Theorem \ref{prop:existence+ktc}}
\label{app:existence+ktc}

\subsubsection{Existence of the Capacity-Achieving Input Distribution}

An optimal distribution exists if the space of input distribution
functions over which the minimization is performed is compact, and
the objective functional is weak* continuous \cite{Luenberger}. The
compactness of the space of input distributions with second moment
constraints is shown in \cite{Abou}. The compactness for the more
stringent case of peak limited inputs follows immediately from this
result. Therefore, we need only to show the weak* continuity of
$I(\cdot|\he)$. The weak* continuity of the functional
$I(\cdot|\he)$ is equivalent to
\begin{gather} \label{eq:equivalentw*cont}
F_{n} \stackrel{w^{*}}{\rightarrow} F \Rightarrow I(F_{n}|\he)
\rightarrow I(F|\he).
\end{gather}
We first note the upper bound
\begin{gather} \label{eq:uppfRr}
f_R(R|r,\K) \le \frac{R^{m-2}}{(m-3)!} \frac{e^{-\frac{R+\K
r^2}{1+r^2}}}{(1+r^2)} I_0\left(\frac{2\sqrt{\K R}\, r}{1+r^2}
\right)
\end{gather}
which is obtained from the bound $(1-a)^{m-3}
e^{\frac{ar^2R}{1+r^2}} \, I_0\left(\frac{2\sqrt{\K R}\,r
\sqrt{a}}{1+r^2} \right) \le e^{\frac{r^2R}{1+r^2}} \,
I_0\left(\frac{2\sqrt{\K R}\,r}{1+r^2} \right)$ $\forall a \in
[0,1]$. The upper bound in (\ref{eq:uppfRr}) is bounded for all $r
\in [0,\sqrt{L}]$ and also for all $R \ge 0$ due to the exponential
decrease in $R$ in the second term. Since $f_R(R|r,\K)$ and
$\log(1+r^2)$ are continuous and bounded functions for all
$r\in[0,\sqrt{L}]$ and $R \ge 0$, by the definition of weak
convergence \cite{Luenberger},
\begin{gather}
F_n \stackrel{w^{*}}{\rightarrow} F \Rightarrow \int_0^\infty
\log(1+r^2) \, \ud F_n(r) \rightarrow \int_0^\infty \log(1+r^2) \,
\ud F(r) \label{eq:weakcontlog}
\\ \intertext{and}
F_n \stackrel{w^{*}}{\rightarrow} F \Rightarrow \int_0^\infty
f_R(R|r,\K) \, \ud F_n(r) \rightarrow \int_0^\infty f_R(R|r,\K) \,
\ud F(r)
\end{gather}
for all $R \ge 0$. Therefore, we have
\begin{gather} \label{eq:weakcontg}
F_n \stackrel{w^{*}}{\rightarrow} F \Rightarrow g(R,F_n,r)
\rightarrow g(R,F,r) \quad \forall R \ge 0.
\end{gather}
Note that the mutual information in (\ref{eq:mutualinfoisotrop}) can
also be written as
\begin{align} \label{eq:mutualinfoisotropapp}
I(F_r | \he) = -\int_0^\infty  d\K \, f_\K(\K)\int_0^\infty dR \,
\frac{R^{m-2}}{(m-2)!}\,
 g(R,F_r, \K ) \log \,g(R,F_r, \K )   - E_r \{\log(1+r^2)\} - (m-1)
\end{align}
The weak* continuity of the second term on the right-hand-side of
(\ref{eq:mutualinfoisotropapp}) follows from (\ref{eq:weakcontlog}).
In order to show (\ref{eq:equivalentw*cont}) and hence the weak*
continuity of the mutual information, we need to prove
\begin{align}
\lim_{n \to \infty} &\int_0^\infty  d\K \, f_\K(\K)\int_0^\infty dR
\, \frac{R^{m-2}}{(m-2)!}\,
 g(R,F_n, \K ) \log \,g(R,F_n, \K ) \label{eq:weakcontint1}
 \\
&= \int_0^\infty  \lim_{n \to \infty} d\K \, f_\K(\K)\int_0^\infty
dR \, \frac{R^{m-2}}{(m-2)!}\,
 g(R,F_n, \K ) \log \,g(R,F_n, \K ) \label{eq:weakcontint2}
 \\
 &= \int_0^\infty  d\K \, f_\K(\K)\int_0^\infty
\lim_{n \to \infty} dR \, \frac{R^{m-2}}{(m-2)!}\,
 g(R,F_n, \K ) \log \,g(R,F_n, \K ) \label{eq:weakcontint3}
 \\
 &= \int_0^\infty  d\K \, f_\K(\K)\int_0^\infty
dR \, \frac{R^{m-2}}{(m-2)!}\,
 g(R,F, \K ) \log \,g(R,F, \K ) \label{eq:weakcontint4}
\end{align}
(\ref{eq:weakcontint4}) follows from (\ref{eq:weakcontg}) and the
continuity of the function $x\log x$. In order to justify the
interchanges of the limit and integral in (\ref{eq:weakcontint2})
and (\ref{eq:weakcontint3}), we invoke the Dominated Convergence
Theorem \cite{Rudin} which requires an integrable upper bound on the
integrand. We first find the following upper bound on the function
$g$:
\begin{align}
g(R,F_n, \K ) &\le (m-2) \int_0^{\sqrt{L}} \frac{e^{-\frac{R+\K
r^2}{1+r^2}}}{(1+r^2)} I_0\left(\frac{2\sqrt{\K R}\,r}{1+r^2}
\right) \, dF_n(r) \label{eq:uppg1}
\\
&\le (m-2) e^{-\frac{R}{1+L} + \sqrt{\K R}} \int_0^{\sqrt{L}}
\frac{e^{-\frac{\K r^2}{1+r^2}}}{(1+r^2)} \, dF_n(r)
\label{eq:uppg2}
\\
&\le (m-2)\, e^{-\frac{R}{1+L} + \sqrt{\K R}} \label{eq:uppg3}
\\
&\triangleq u(R,\K) \label{eq:uppg4} \quad \forall n, \quad \forall
R,\K \ge 0.
\end{align}
(\ref{eq:uppg1}) follows from the upper bound in (\ref{eq:uppfRr}).
(\ref{eq:uppg2}) is obtained by noting that $e^{-\frac{R}{1+r^2}}
\le e^{-\frac{R}{1+L}}$ for all $r \in [0,\sqrt{L}]$ and $R \ge 0$,
and $I_0\left(\frac{2\sqrt{\K R}\, r}{1+r^2} \right) \le
I_0(\sqrt{\K R}) \le e^{\sqrt{\K R}}$ $\forall R,r \ge 0$. Finally,
(\ref{eq:uppg3}) follows from the observation that the integrand in
(\ref{eq:uppg2}) is less than 1 $\forall r,\K \ge 0$. Note that the
upper bound $u(R,\K)$ is not a function of $F_n$ and decreases
exponentially in $R$ for sufficiently large values of $R$. Next, we
find the following upper bound:
\begin{align}
\left| \frac{R^{m-2}}{(m-2)!}\,
 g(R,F_n, \K ) \log \,g(R,F_n, \K ) \right| &\le
 \frac{R^{m-2}}{(m-2)!}\, (4 \,g^{0.9}(R,F_n, \K ) + g^2(R,F_n, \K
 )) \label{eq:uppinteg1}
\\
&\le \frac{R^{m-2}}{(m-2)!}\, (4 \,u^{0.9}(R,\K ) + u^2(R,\K ))
\quad \forall R,\K \ge 0. \label{eq:uppinteg2}
\end{align}
(\ref{eq:uppinteg1}) follows from the fact that $|x \log(x)| \le
4x^{0.9} + x^2$ for all $x \ge 0$, and (\ref{eq:uppinteg2}) follows
from (\ref{eq:uppg4}). Note that the upper bound in
(\ref{eq:uppinteg2}) does not depend on $F_n$ and is integrable due
to the exponential decay of $u(R,\K)$ in $R$ for sufficiently large
values of $R$. Applying the Dominated Convergence Theorem with the
upper bound in (\ref{eq:uppinteg2}) justifies
(\ref{eq:weakcontint3}). We further consider
\begin{align}
\left| f_\K(\K)\!\!\int_0^\infty \!\!\!\!\!dR \,
\frac{R^{m-2}}{(m-2)!}\,
 g(R,F_n, \K ) \log \,g(R,F_n, \K )\right| &\le f_\K(\K)\int_0^\infty \!\!\!\!dR \, \frac{R^{m-2}}{(m-2)!}
 \left|g(R,F_n, \K ) \log \,g(R,F_n, \K )\right|
 \\
&\le f_\K(\K)\int_0^\infty \!\!dR \, \frac{R^{m-2}}{(m-2)!}\,
 (4 \,g^{0.9}(R,F_n,\K ) + g^2(R,F_n,\K )) \label{eq:uppintegfK1}
\end{align}
Note that $f_\K(\K) = \frac{1}{E\{\K\}} e^{-\frac{\K}{E\{\K\}}}$
where $E\{\K\} = \frac{\gamma^2 \delta m P}{N_0}$. The integral of
the upper bound $u(R,\K)$ with respect to $R$ increases
exponentially with $\K$. Hence, we need to find a tighter upper
bound. We have
\begin{align}
g(R,F_n, \K ) &\le (m-2) \int_0^{\sqrt{L}} \frac{e^{-\frac{R+\K
r^2}{1+r^2}}}{(1+r^2)} I_0\left(\frac{2\sqrt{\K R}\,r}{1+r^2}
\right) \, dF_n(r) \label{eq:uppg2_1}
\\
&\le (m-2) \int_0^{\sqrt{L}} e^{-\frac{R+\K r^2 - 2\sqrt{\K
R}\,r}{1+r^2}} \, dF_n(r) \label{eq:uppg2_2}
\\
&\le (m-2) \int_0^{\sqrt{L}} e^{-\frac{(\sqrt{R}-\sqrt{\K
}\,r)^2}{1+L}} \, dF_n(r) \label{eq:uppg2_3}
\\
&\le \left\{
\begin{array}{ll}
(m-2) & R \le \K L
\\
(m-2) e^{-\frac{(\sqrt{R}-\sqrt{\K L})^2}{1+L}} & R > \K L
\end{array} \right. \label{eq:uppg2_4}
\\
&\triangleq v(R,\K) \label{eq:uppg2_5} \quad \forall n, \quad
\forall R,\K \ge 0
\end{align}
where (\ref{eq:uppg2_2}) follows from the fact that $I_0(x) \le
e^{x}$, and (\ref{eq:uppg2_3}) follows by choosing the largest value
$r = \sqrt{L}$ in the denominator of the exponential function.
(\ref{eq:uppg2_4}) is obtained by noting that $(\sqrt{R}-\sqrt{\K
}\,r)^2$ is a nonnegative quadratic function of $r$, minimized at $r
= \sqrt{\frac{R}{K}}$. Hence, if $L \ge \frac{R}{K}$, the minimum
value of the quadratic function is zero. Otherwise, it is
$(\sqrt{R}-\sqrt{\K L})^2$. From (\ref{eq:uppintegfK1}) and
(\ref{eq:uppg2_5}), we have
\begin{align}
\left| f_\K(\K)\!\!\int_0^\infty \!\!\!\!\!dR \,
\frac{R^{m-2}}{(m-2)!}\,
 g(R,F_n, \K ) \log \,g(R,F_n, \K )\right|
&\le f_\K(\K)\int_0^\infty \!\!dR \, \frac{R^{m-2}}{(m-2)!}\,
 (4 \,v^{0.9}(R,\K ) + v^2(R,\K )). \label{eq:uppintegfK2}
\end{align}
Note that the upper bound in (\ref{eq:uppintegfK2}) is independent
of $F_n$. It can also be verified easily that this upper bound is
integrable with respect to $\K$ due to the facts that $f_\K$
decreases exponentially with $\K$ while the integral in the upper
bound produces a result that is at most polynomial in $\K$. Applying
the Dominated Convergence Theorem with the integrable upper bound in
(\ref{eq:uppintegfK2}) justifies (\ref{eq:weakcontint2}). Hence, the
proof is complete.
\\

\subsubsection{Sufficient and Necessary Kuhn-Tucker Condition}

The proof of the sufficient and necessary condition in
(\ref{eq:ktc}) follows along the same lines as those in \cite{Abou}
and \cite{Gursoy-part1}. The weak derivative of $I(\cdot| \he)$ at
$F_0$ is defined as
\begin{align}
I_{F_0}^{'}(F|\he) \triangleq \lim_{\theta \to 0}
\frac{I[(1-\theta)F_0 + \theta F|\he]-I(F_0|\he)}{\theta}.
\end{align}
The weak derivative of the mutual information in
(\ref{eq:mutualinfoisotrop}) is obtained as
\begin{align}
&I_{F_0}^{'}(F|\he) = E_{\K} \left\{ \int dF_0(r) \int_0^\infty
f_{R|r,\K}(R|r,\K) \log \,g(R,F_0, \K ) \, dR \right\} \nonumber
\\
&- E_{\K} \left\{ \int dF(r) \int_0^\infty f_{R|r,\K}(R|r,\K) \log
\,g(R,F_0, \K ) \, dR \right\} + \int dF_0(r) \log(1+r^2) - \int
dF(r) \log(1+r^2).
\end{align}
Note that if $F_0$ is indeed the maximizing distribution and hence
capacity achieving, then $I_{F_0}^{'}(F|\he) \le 0$ for all $F$
satisfying the peak power constraint. Then using the same steps in
\cite[Appendix II, Theorem 4]{Abou}, we can show that $F_0$ is a
capacity-achieving input distribution if and only if
\begin{align}
E_{\K} \left\{ \int_0^\infty f_{R|r,\K}(R|r,\K) \log \,g(R,F_0, \K )
\, dR\right\} + \log(1+r^2) + mC_\delta + m-1 \ge 0 \quad \forall r
\in [0,\sqrt{L}]
\end{align}
with equality at the points of increase of distribution $F_0$.

\subsection{Analyticity of the Kuhn-Tucker Condition in the Complex
Domain} \label{app:analyticity}

We consider the following function which is the left-hand-side of
the Kuhn-Tucker condition (\ref{eq:ktc}) in the complex domain:
\begin{gather} \label{eq:ktccomplexapp}
\Phi(z) = E_{\K} \left\{ \int_0^\infty f_{R|r,\K}(R|z,\K) \log
\,g(R,F_r, \K ) \, dR \right\} + \log(1+z) + mC_\delta + (m-1).
\end{gather}
Note that $\log(1+z)$ is an analytic function of $z = z_r + jz_i$ in
the entire complex plane excluding the real axis with $z_r \le -1$
because the principle branch of the logarithm is not analytic only
on the negative real line. Next, we investigate the region in which
the first term of (\ref{eq:ktccomplexapp}) is analytic. We first
note the Differentiation Lemma.

\begin{lemma:differentiation}\cite[Sec. XII]{Lang}
Let $I$ be an interval of real numbers, possibly infinite. Let $U$
be an open set of complex numbers. Let $f = f(t,z)$ be a continuous
function on $I \times U$. Assume:

(i) For each compact subset $K$ of $U$ the integral $\int_I f(t,z)
\, \ud t$ is uniformly convergent for $s \in K$.

(ii) For each $t$ the function $z \mapsto f(t,z)$ is analytic. Let
$F(z) = \int_I f(t,z) \, \ud t$.
\\
Then $F$ is analytic on $U$ and $F^{'}(z) = \int_I D f(t,z) \, \ud
t$ where $D$ is the differentiation operator. Furthermore $D f(t,z)$
satisfies the same hypothesis as $f$. \hfill $\square$
\end{lemma:differentiation}

The integral $\int_0^{\infty} f(t,z) \, \ud t$ is said to be
uniformly convergent \cite{Lang} for $z \in K$ if, given $\epsilon >
0$, there exists $B_0$ such that if $B_0 < B_1 < B_2$, then
$\left|\int_{B_1}^{B_2}f(t,z) \, \ud t \right| < \epsilon$. From
this definition it can be easily shown that if $\int_0^\infty
|f(t,z)| \, dt < \infty$, then $\int_0^{\infty} f(t,z) \, \ud t$  is
uniformly convergent.

The function

\begin{gather}
f_{R|r,\K}(R|z,\K) = \frac{R^{m-2}}{(m-3)!} \, \frac{e^{-R -
\frac{\K z^2}{1+z^2}}}{1+z^2} \int_0^1 (1-a)^{m-3}
e^{\frac{az^2R}{1+z^2}} \, I_0\left(\frac{2\sqrt{\K R}\,
z\sqrt{a}}{1+z^2} \right) \, da \label{eq:fRzapp}
\end{gather}
is analytic in the entire complex plane excluding the points at $z =
\pm j$ because rational functions are analytic everywhere except at
the points that make the denominator zero; the exponential function
and $I_0$ are analytic everywhere because they can be expanded as
power series; and if $g$ and $f$ are analytic then $g \circ f$ is
also analytic in the corresponding region. The analyticity of the
integral in (\ref{eq:fRzapp}) can also be easily verified using the
Differentiation Lemma since the integration is over a finite
interval. In order to find the region in which the the first term on
the right-hand-side of (\ref{eq:ktccomplexapp}) is analytic, we need
to find the region $\D$ that satisfies for all $z \in \D $
\begin{gather} \label{eq:intfRzapp}
\int_0^\infty f_\K(\K) \int_0^\infty |f_{R|r,\K}(R|z,\K)| |\log
\,g(R,F_r, \K )| \, dR < \infty.
\end{gather}
We consider
\begin{align}
|f_{R|r,\K}(R|z,\K)| &= \left|\frac{R^{m-2}}{(m-3)!} \, \frac{e^{-
\frac{\K z^2}{1+z^2}}}{1+z^2} \int_0^1 (1-a)^{m-3} e^{-R \frac{1 +
(1-a)z^2}{1+z^2}} \, I_0\left(\frac{2\sqrt{\K R}\, z\sqrt{a}}{1+z^2}
\right) \, da\right| \label{eq:fRzabs1}
\\
&\le \frac{R^{m-2}}{(m-3)!} \, \frac{\left|e^{-\frac{\K
z^2}{1+z^2}}\right|}{|1+z^2|} \int_0^1 (1-a)^{m-3} \left|e^{-R
\frac{1 + (1-a)z^2}{1+z^2}}\right| \, \left|I_0\left(\frac{2\sqrt{\K
R}\, z\sqrt{a}}{1+z^2}\right)\right| \, da \label{eq:fRzabs2}
\\
&\le \frac{R^{m-2}}{(m-3)!} \, \frac{e^{-\Re\left\{\frac{\K
z^2}{1+z^2}\right\}}}{|1+z^2|} \int_0^1 (1-a)^{m-3} e^{-R \,
\Re\left\{\frac{1 + (1-a)z^2}{1+z^2}\right\}} \, I_0\left(2\sqrt{\K
R}\sqrt{a}\,\Re\left\{\frac{z}{1+z^2}\right\}\right) \, da
\label{eq:fRzabs3}
\\
&\le \frac{R^{m-2}}{(m-3)!} \, \frac{e^{-\Re\left\{\frac{\K
z^2}{1+z^2}\right\}}}{|1+z^2|} \int_0^1 (1-a)^{m-3} e^{-R \,
\Re\left\{\frac{1 + (1-a)z^2}{1+z^2}\right\}} \, e^{2\sqrt{\K
R}\sqrt{a}\,\left|\Re\left\{\frac{z}{1+z^2}\right\}\right|} \, da
\label{eq:fRzabs4}
\\
&\le \frac{R^{m-2}}{(m-3)!} \, \frac{e^{-\Re\left\{\frac{\K
z^2}{1+z^2}\right\}}}{|1+z^2|} \int_0^1 (1-a)^{m-3} e^{-R \,
\frac{1+z_r^2 - z_i^2}{|1+z^2|^2}} \, e^{2\sqrt{\K
R}\,\left|\Re\left\{\frac{z}{1+z^2}\right\}\right|} \, da
\label{eq:fRzabs5}
\\
&= \frac{R^{m-2}}{(m-2)!} \, \frac{e^{-\Re\left\{\frac{\K
z^2}{1+z^2}\right\}}}{|1+z^2|} \, e^{-R \, \frac{1+z_r^2 -
z_i^2}{|1+z^2|^2}} \, e^{2\sqrt{\K
R}\,\left|\Re\left\{\frac{z}{1+z^2}\right\}\right|}
\label{eq:fRzabs6}
\\
&= \frac{R^{m-2}}{(m-2)!} \, \frac{e^{-\K\frac{(z_r^2 -
z_i^2)(1+z_r^2-z_i^2)+4z_r^2z_i^2}{|1+z^2|^2}}}{|1+z^2|} \, e^{-R \,
\frac{1+z_r^2 - z_i^2}{|1+z^2|^2}} \, e^{2\sqrt{\K
R}\,\frac{|z_r(1+z_r^2-z_i^2) +2z_rz_i^2|}{|1+z^2|^2}}
\label{eq:fRzabs7}
\\
&= \frac{R^{m-2}}{(m-2)!} \, \frac{e^{-\K\frac{(z_r^2 -
z_i^2)(1+z_r^2-z_i^2)+4z_r^2z_i^2}{|1+z^2|^2}}}{|1+z^2|} \,
e^{-\frac{\left(\sqrt{R(1+z_r^2 - z_i^2)} -
\sqrt{\K}\,\frac{|z_r(1+z_r^2-z_i^2) +2z_rz_i^2|}{\sqrt{1+z_r^2 -
z_i^2}}\right)^2}{|1+z^2|^2}} \, e^{\K \frac{(z_r(1+z_r^2-z_i^2)
+2z_rz_i^2)^2}{(1+z_r^2 - z_i^2)|1+z^2|^2}} \label{eq:fRzabs8}
\\
&= \frac{R^{m-2}}{(m-2)!} \, \frac{e^{-\K\left(\frac{(z_r^2 -
z_i^2)(1+z_r^2-z_i^2)+4z_r^2z_i^2}{|1+z^2|^2} -
\frac{(z_r(1+z_r^2-z_i^2) +2z_rz_i^2)^2}{(1+z_r^2 -
z_i^2)|1+z^2|^2}\right)}}{|1+z^2|} \,
e^{-\frac{\left(\sqrt{R(1+z_r^2 - z_i^2)} -
\sqrt{\K}\,\frac{|z_r(1+z_r^2-z_i^2) +2z_rz_i^2|}{\sqrt{1+z_r^2 -
z_i^2}}\right)^2}{|1+z^2|^2}} \label{eq:fRzabs9}
\\
&= \frac{R^{m-2}}{(m-2)!} \, \frac{e^{\K\left(\frac{
z_i^2(1+z_r^2-z_i^2)+\frac{4z_r^2z_i^4}{1+z_r^2-z_i^2}}{|1+z^2|^2}\right)}}{|1+z^2|}
\, e^{-\frac{\left(\sqrt{R(1+z_r^2 - z_i^2)} -
\sqrt{\K}\,\frac{|z_r(1+z_r^2-z_i^2) +2z_rz_i^2|}{\sqrt{1+z_r^2 -
z_i^2}}\right)^2}{|1+z^2|^2}} \label{eq:fRzabs10}
\end{align}
In the above formulations, $\Re(z)$ denotes the real value of the
complex-valued number $z = z_r + j z_i$ whose real and imaginary
components are also denoted by $z_r$ and $z_i$, respectively.
(\ref{eq:fRzabs2}) follows by taking the absolute value of the
integrand instead of the absolute value of the integral.
(\ref{eq:fRzabs3}) follows from the facts that $\Re\left(
e^{z}\right) = e^{\Re(z)}$ and $|I_0(z)| \le I_0(\Re(z))$.
(\ref{eq:fRzabs4}) is due to $I_0(x) \le e^{|x|}$ for a real number
$x$. (\ref{eq:fRzabs5}) is obtained from the bounds
$\Re\left\{\frac{1 + (1-a)z^2}{1+z^2}\right\} \ge \frac{1+z_r^2 -
z_i^2}{|1+z^2|^2}$ which holds for all $a \in [0,1]$ and $|z_r| \ge
|z_i|$, and $e^{2\sqrt{\K
R}\sqrt{a}\,\left|\Re\left\{\frac{z}{1+z^2}\right\}\right|} \le
e^{2\sqrt{\K R}\,\left|\Re\left\{\frac{z}{1+z^2}\right\}\right|}$
$\forall a \in [0,1]$. (\ref{eq:fRzabs6}) follows by evaluating the
integral in (\ref{eq:fRzabs5}), in which the only term that depends
on $a$ is $(1-a)^{m-3}$. (\ref{eq:fRzabs7}) is obtained by
explicitly expressing $\Re\left\{\frac{\K z^2}{1+z^2}\right\}$ and
$\Re\left\{\frac{z}{1+z^2}\right\}$ in terms of $z_r$ and $z_i$, the
real and imaginary components of $z$. (\ref{eq:fRzabs8}) follows by
expressing the exponents of the second and third exponential
functions as a quadratic function of $\sqrt{R}$. Eventually,
(\ref{eq:fRzabs10}) is obtained from straightforward algebraic
computations.

The following lower bound on $g(R,F_r,\K)$ can easily be verified by
noting that $e^{x} \ge 1$ and $I_0(x) \ge 1$  for all $x \ge 0$:
\begin{align}
g(R,F_r,\K) \ge e^{-R} \int_{0}^{L} \frac{e^{-\frac{\K
r^2}{1+r^2}}}{1+r^2} \, dF_r \ge e^{-R} \, \frac{e^{-\frac{\K
L}{1+L}}}{1+L}.
\end{align}
From the above lower bound, we see that $|\log g(R,F_r,\K)|$
increases at most linearly in both $R$ and $\K$ for sufficiently
large values of $R$ and $\K$. Therefore, if $(1+z_r^2 - z_i^2) > 0$,
then the upper bound in (\ref{eq:fRzabs10}) decreases exponentially
in $R$, and as a result, the inner integral in (\ref{eq:intfRzapp})
converges. This condition is satisfied in the region where $|z_r|
\ge |z_i|$.

Note that the upper bound in (\ref{eq:fRzabs10}) increases
exponentially in $\K$. However, the value of the function
\begin{gather}
c(z_r,z_i) =
\frac{z_i^2(1+z_r^2-z_i^2)+\frac{4z_r^2z_i^4}{1+z_r^2-z_i^2}}{|1+z^2|^2}
\end{gather}
can be made arbitrarily small by choosing arbitrarily small values
for $|z_i|$. Note also that $f_\K(\K) = \frac{1}{E\{\K\}}
e^{-\frac{\K}{E\{\K\}}}$ where $E\{\K\} = \frac{\gamma^2 \delta m
P}{N_0}$. Hence, in the region where $c(z_r,z_i) <
\frac{N_0}{\gamma^2 \delta m P}$, we have the integrand in
(\ref{eq:intfRzapp}) exponentially decreasing in $\K$ and as a
result the integral converges. it can be shown that for a fixed
$|z_i| < 1$, $c(z_r, z_i)$ is a monotonically decreasing function of
$z_r \ge 0$ achieving its maximum of at $\frac{z_i^2}{1-z_i^2}$ at
$z_r = 0$. Hence, we consider the following region in the complex
domain:
\begin{align}
\D = &\left\{(z_r,z_i): 0 \le z_r \le
\min\left(\frac{1}{\sqrt{2}},\sqrt{\frac{N_0}{2\gamma^2 \delta m
P}}\right) \text{ and }|z_i| \le z_r \right\} \nonumber
\\
& \bigcup \,\left\{(z_r,z_i): z_r >
\min\left(\frac{1}{\sqrt{2}},\sqrt{\frac{N_0}{2\gamma^2 \delta m
P}}\right) \text{ and }|z_i| \le
\min\left(\frac{1}{\sqrt{2}},\sqrt{\frac{N_0}{2\gamma^2 \delta m
P}}\right) \right\}
\end{align}
In region $\D$, $c(z_r,z_i) < \frac{N_0}{\gamma^2 \delta m P}$ and
$|z_r| \ge |z_i|$. Hence, the integral in (\ref{eq:intfRzapp})
converges in this region. Moreover, this region includes the
positive real line.

\end{document}